\documentclass[final,5p,times,authoryear]{elsarticle}
\usepackage[utf8]{inputenc}

\usepackage{amssymb,hyperref}
\usepackage{lipsum}

\journal{Astronomy $\&$ Computing}
\def\apjs{{\it Astrophys.~J.~Suppl.}}

\def\apjl{{\it Astrophys.~J.~Lett.}}

\def\aap{{\it Astron.~Astrophys.}}

\DeclareUnicodeCharacter{2212}{-}
\begin{document}

\begin{frontmatter}



\title{Membership determination in open clusters using the DBSCAN Clustering Algorithm}

\author[second]{Mudasir Raja}
\affiliation[second]{organization={Maulana Azad National Urdu University},
            addressline={Gachibowli}, 
            city={Hyderabad},
            postcode={500 032},              state={Telangana},
            country={India}} 
\author[second, third]{Priya Hasan}
\affiliation[third]{organization={Inter-University Centre for Astronomy and Astrophysics},
            addressline={Post Bag 4
Ganeshkhind,
Savitribai Phule Pune University Campus}, 
            city={Pune},
            postcode={411 007},              state={Maharashtra},
            country={India}} 

\author[first]{Md Mahmudunnobe}
\affiliation[first]{organization={Wayne State University},
            addressline={42 W Warren Ave}, 
            city={ Detroit},
            postcode={48202},              state={MI},
            country={USA}}

\author[second]{Md Saifuddin}
\author[second]{S N Hasan}

\begin{abstract}
In this paper, we  apply the machine learning clustering algorithm Density 
 Based Spatial Clustering of Applications with Noise (DBSCAN) to study the membership of stars in twelve open clusters (NGC~2264, NGC~2682, NGC~2244, NGC~3293, NGC~6913, NGC~7142, IC~1805, NGC~6231, NGC~2243, NGC 6451, 
NGC 6005 and NGC 6583) based on Gaia DR3 Data. 
This sample of clusters spans a variety of parameters like age, metallicity, distance, extinction and a wide parameter space in proper motions and parallaxes. We  obtain reliable cluster members using DBSCAN as faint as $G \sim 20$ mag and also in the outer regions of clusters. With our revised membership list, we plot color-magnitude diagrams and we obtain cluster
parameters for our sample using ASteCA and compare it with the catalog values. We also validate our membership sample by spectroscopic data from APOGEE and GALAH for the available data.  This paper demonstrates the effectiveness of DBSCAN in membership determination of clusters.  
\end{abstract}



\begin{keyword}
(Galaxy:) open clusters and associations: general, individual: -- (stars:) Hertzsprung–Russell and color-magnitude diagrams--machine learning: --DBSCAN


\end{keyword}

\end{frontmatter}




\section{Introduction} \label{sec1}

Star clusters are the fundamental components of galaxies and play a vital role in understanding the formation and evolution of stars and galaxies \citep{janes1994galactic, friel1995old, doi:10.1146/annurev-astro-091918-104430}. An open cluster is a stellar system which is gravitationally bound and is made up of tens to thousands of stars and present an irregular and a loosely bound structure. Stars in a cluster are born from the same molecular cloud, at about the same time and hence they approximately are of the same chemical composition, same age  and at the same distance. Most of the open clusters are located close to the Galactic plane and thus serve as excellent tracers of the recent formation history of the Galactic disc \citep{friel1995old,chen2003galactic,jacobson2016gaia}. 
 Gaia DR3 \citep{prusti2016gaia, collaboration2017gaia, brown2021gaia, hasan2021gaia,collaboration2022gaia} includes precise astrometry at the sub-milliarcsecond level  and broad band photometry for a total of 1.8 billion objects based on 34 months of satellite data. 

Previously, due to non-availability of extensive data on stars, many researchers studied specific regions (which often contained a small central region of an open star cluster), and  employed membership determination methods that had limited application. The popular method for  member determination of open clusters was the use of proper-motion data by the Vasilevskis-Sanders method
(also called the VS method) proposed by 
 \cite{vasilevskis1958relative} which was improved by \cite{sanders1971improved}.  \cite{zhao1990improved} proposed an improved scheme to enable use of the VS method with the proper motion data of unequal inaccuracies.
Other methods include \citep{cabrera1990non, galadi1998overlapping,castro2018new}. Photometric methods have also been used for membership determination\citep{Hasan_2011}. 

Machine Learning (ML) provides improved methods of membership determination. Random Forest (RF)
is a supervised classification method and was applied to the Gaia DR2 data in \citep{gao2018machine,mahmudunnobe2021membership}. Guassian Mixture Modeling (GMM) was also explored by us \cite{mahmudunnobe2024using}.
\cite{2020A&A...640A...1C} obtained reliable parameters for 1867 clusters  by training an artificial neural network to estimate parameters from the Gaia DR2 photometry of cluster members and their mean parallax.
\cite{gao2014membership} proposed to use a classical algorithm in the data mining—
Density-Based Spatial Clustering of Applications with Noise (DBSCAN) clustering
algorithm \citep{ester1996proc} for the determination of members in open star clusters using the fact that cluster stars would be `clustered' not only spatially, but also in parameter space. By using the 3D kinematic data (two-dimensional proper motion and radial velocity)  of 1046 stars \cite{gao2014membership} determined members of NGC~188 and validated it by plotting the color-magnitude diagram.
\cite{2018A&A...618A..59C}  used DBSCAN to find clusters in the Gaia DR2 data coupled with a supervised learning method such as an artificial neural network (ANN) to  distinguish between real OCs and statistical clusters in an automated method.
\cite{shou2019dbscan} used the DBSCAN clustering algorithm for the detection of nearby open clusters based on Gaia DR2 within 100 pc of the Sun. \cite{2020A&A...635A..45C}  used DBSCAN to blindly search for these overdensities in Gaia DR2 and then applied a deep learning artificial neural network trained on colour-magnitude diagrams to identify isochrone patterns in these overdensities, and confirm them as open clusters.
\cite{2022A&A...661A.118C} aimed to complete the open cluster census in the Milky Way with the detection of new stellar groups in the Galactic disc using an OCfinder method. \cite{2021RAA....21...93H} used Gaia DR2 to blindly search for open star clusters in the Milky Way within the Galactic latitude range of  $b < 20^0$ using  an unsupervised machine learning method. \cite{2022ApJS..262....7H} searched Gaia EDR3 for nearby (parallax $>$ 0.8 mas) all-sky regions, to find  886 star clusters, of which 270 candidates were new. \cite{2023ApJS..267...34H}  and earlier references by the authors, used DBSCAN algorithm to identify clusters on fainter and more distant stars. 

The DBSCAN algorithm assumes an average density of stars in the region. In the case of regions that are small, the average is a good approximation compared to wider regions of the sky, where the density is not constant. In the case of wider searches, HDBSCAN proves more effective since it applies variable densities as used by \cite{Hunt_2021,2023A&A...673A.114H}.

In an earlier work, we used DBSCAN, OPTICS and HDBSCAN to identify YSOs in Gaia data to study their distribution and kinematics in the Serpens Cloud \cite{2023JApA...44...41H}.
 
As an extension of our exploration of different algorithms of membership determination \citep{mahmudunnobe2021membership,mahmudunnobe2024using, 2023JApA...44...41H}, in this paper, we shall focus on DBSCAN which is an unsupervised density based clustering technique that can identify clusters of non-spherical shapes and hence is highly suitable for open clusters in smaller regions of approximately constant densities. We use DBSCAN for membership determination using Gaia DR3 data \citep{2023A&A...674A...1Gca} for a sample of twelve clusters. We provide the code for use by the community and queries can be addressed  to PH \footnote{The code used for a sample cluster is available at \url{https://github.com/priyahasan/ML/blob/master/DBSCAN_6913.ipynb}}
\subsection{$Epsilon$ and $MinPts$}. 
 
The paper is structured as follows: Section \ref{sec1}  gives a brief introduction and a description of the problem. Section \ref{sec2} describes the data and sample selection. Section \ref{sec3} describes the DBSCAN Method.  Section \ref{sec4} describes the DBSCAN results.  Section \ref{sec5} describes the APOGEE and GALAH data. Section \ref{sec6} presents ASteCA results and Section \ref{sec7} is the Discussions and Conclusions part of our paper.
 
\section{Cluster Sample}\label{sec2}
In this section we determine the membership of twelve open clusters: NGC~2264, NGC~2682, NGC~2244, NGC~3293, NGC~6913, NGC~7142, IC~1805, NGC~6231, NGC~2243, NGC 6451, NGC 6005 and NGC 6583. 

\begin{figure}
\includegraphics[width=0.45\textwidth]{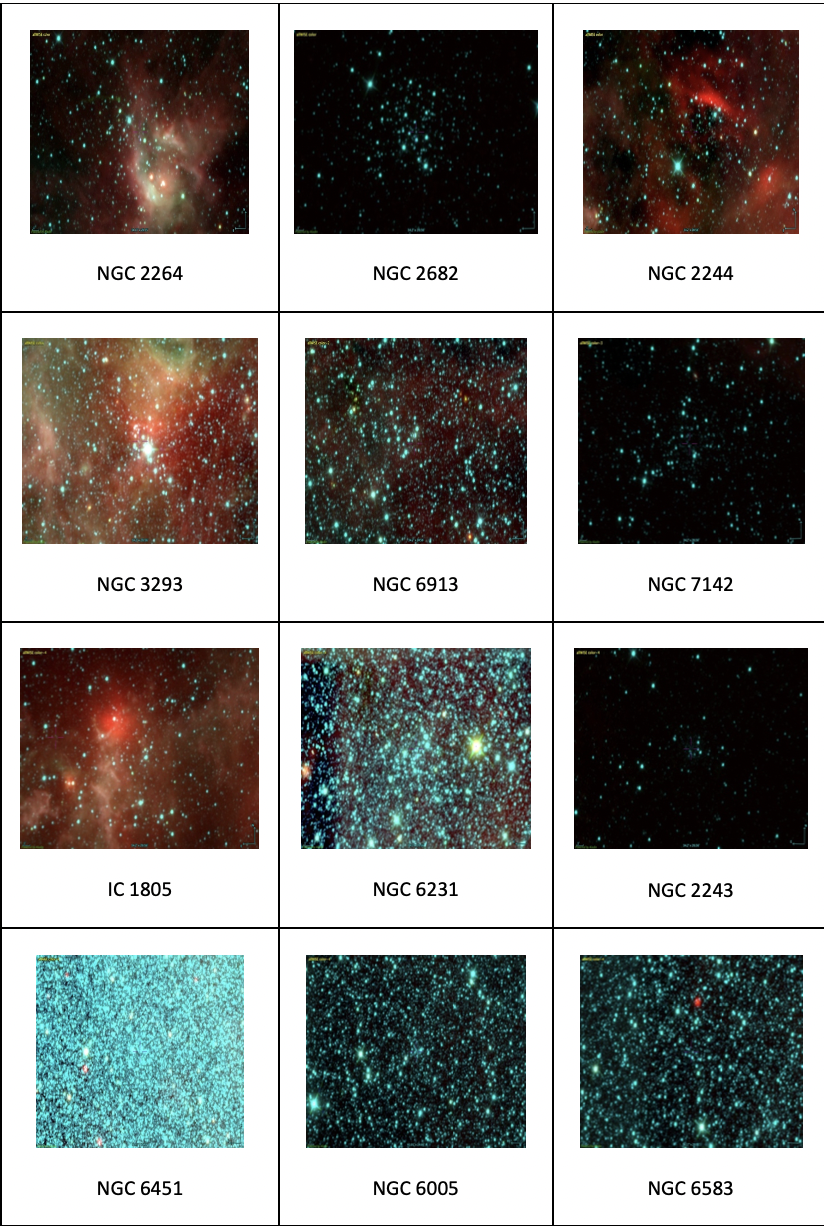}
 \caption {WISE images of the cluster sample: NGC~2264, NGC~2682, NGC~2244, NGC~3293, NGC~6913, NGC~7142, IC~1805, NGC~6231, NGC~2243, NGC 6451, NGC 6005 and NGC 6583. In all images, North is up, East is left. Field of View: '34.2 X 20.6'}
 \label{param}
 \end{figure}

 These clusters have distances ranging from 707~pc to 3719~pc. 
The basic parameters of the selected clusters are given in Table \ref{tab:clusterdata} which shows the coordinates of these clusters ($\alpha$ and $\delta$), the angular diameter ($r50$), the radius that contains half the number of members from the same reference, the logarithm of age $log \ t$, the extinction $A_V$, the distance to the cluster in parsecs is $d$ and $GC$ is the galactocentric distance in parsecs from  \citep{cantat2020clusters}. The WISE images of the cluster sample are shown in Fig. \ref{param}. We choose to present these as WISE images are in the infrared and can give an idea of the reddening (gas and dust) in the cluster sample. This is a good indicator of the possible obscured stars since Gaia observes in the optical.

\begin{table}
\small
\caption[Basic cluster parameters]{Basic cluster parameters \citep{cantat2020clusters}}
\label{tab:clusterdata}
\begin{tabular}{llllllll}
\hline
  Cluster &$\alpha$ & $\delta$ &$r50$ & log t & $A_v$ & $d$ &  $GC$   \\ 
 &  (deg) & (deg) &  (deg) &   & mag & (pc) &  (pc)   \\  
\hline
\hline
\vspace{0.5cm}
NGC 2264 & 100.2 & 9.8 & 0.072 & 7.44 & 0.79 & 707 & 8995\\
\vspace{0.5cm}
NGC 2682  & 132.85 & 11.8 & 0.166 & 9.63 & 0.07 & 899 & 8964 \\ 
\vspace{0.5cm}
NGC 2244 & 98.05 & 4.9 & 0.232 & 7.1 & 1.46 & 1478 & 9686 \\
\vspace{0.5cm}
NGC 3293 & 157.97 & -58.23 & 0.066 & 7.01 & 0.9 & 2710 & 8034 \\
\vspace{0.5cm}
NGC 6913 & 305.9 & 38.48 & 0.106 & 7.34 & 1.8 & 1608 & 8126 \\
\vspace{0.5cm}
NGC 7142 & 326.29 & 65.78 & 0.102 & 9.49 & 1.16 & 2406 & 9255 \\
\vspace{0.5cm}
IC 1805  & 38.21  & 61.47 & 0.11 & 6.88 & 2.22 & 1964 & 9821 \\
\vspace{0.5cm}
NGC 6231 &  253.545 & -41.812 & 0.15 & 7.14& 1.07& 1475& 6938\\
\vspace{0.5cm}
NGC 2243  & 97.4   & -31.28&0.046 & 9.64 & 0.02 &3719 & 10584 \\
\vspace{0.5cm}
NGC 6451 & 267.675 & -30.206  & 0.078 & 7.41 & 2.2 & 2777.0 & 5563 \\
\vspace{0.5cm}
NGC 6005 & 238.955 & -57.439 & 0.057 & 9.1 & 1.42 & 2383.0 & 6510 \\ 
\vspace{0.5cm}
NGC 6583 & 273.962 & -22.143 & 0.046 & 9.08 & 1.52 & 2053.0 & 6323 \\
\hline
\end{tabular}
\end{table}

\section{DBSCAN Analysis}\label{sec3}
For each cluster, we used a search radius which was double the maximum radius found in \cite{cantat2020clusters}, where the maximum radius is defined as the distance of the farthest members from the center. 
Henceforth \cite{cantat2020clusters} members will be referred to as CG members.
The stars extracted from the Gaia archive were filtered to ensure quality check where the parallax is greater than zero, parallax over error was greater than or equal to 3 and the errors in proper motion in ra and dec were less than 1. Also to ensure single stars, the RUWE factor was taken to be less than 1.4. 

Before applying DBSCAN, we first normalized the data to ensure that the difference in the range of the variables does not influence the analysis. 
The data was normalized using the Z-score normalization, where we used median instead of mean, since mean is affected strongly by outliers. We therefore subtracted the median of the feature from each value, and then divided it by the standard deviation, resulting in values that have a median of 0 and a standard deviation of 1.

The DBSCAN model defines groups as dense regions in parameter space that are separated by the sparser regions. It has the advantage over GMM of identifying non-symmetric regions.  As the data was queried in RA and Dec, the DBSCAN was run for pmra, pmdec and parallax. \textit{MinPts} represents the minimum number of points required to form a group, whereas epsilon ($\epsilon$) is the radius around each point where DBSCAN searches for neighboring points.
DBSCAN model starts with a random point and finds all the points within its $\epsilon$-neighborhood. The distance in parameter space between two points $i$ and $j$ is defined as:

$$d_{ij} = \sqrt{ \sum_{n=1}^{k} (x_{i,n} - x_{j,n})^2}$$
\noindent
where $k$ is the number of dimensions of our data.
If a point has atleast $MinPts$ in its $\epsilon$-neighborhood, it is considered a core point.
If the number of neighbors is less than $MinPts$, but the point itself is a neighbor of a core point, it is considered a border point.  The points which are neither core nor border, are labeled as noise points.
The neighboring core and border points form a cluster. After running DBSCAN, we get one or more clusters of stars along with noise points. Finally, we chose the DBSCAN cluster with the larger number of stars as the members of the cluster and the noise points as the field stars.  We may get one or more clusters, but the user has to decide which is the cluster in consideration.

\subsection{Model Parameter Optimization}

Like any machine learning algorithm, the model parameters  $\epsilon$ and \textit{Minpts} 
directly affect the clustering effectiveness and therefore we need to optimize these parameters to get the most efficient model.
Increasing \textit{Minpts} makes the algorithm more conservative, as it requires denser regions to form a group, potentially ignoring smaller, less dense groups.
Similarly smaller $\epsilon$ results in compact groups, while a larger $\epsilon$ may merge different groups and/or noise.

To get the most optimal value for these two parameters, we used elbow plots or $k$-dist plots as used by  \cite{shou2019dbscan, ester1996kdd} and mean nearest neighbors were found as in \cite{gao2014membership}. 
For our analysis, we used a combination of $k$-dist plot, Mean Nearest Neighbour ($MNN$), Modified Silhouette Score ($MSS$) and number of retrieved members to optimize $\epsilon$ and $MinPts$ which are discussed below.

\subsubsection{$k$-dist Plots:}

Using this principle, if we find a clear elbow point in the $k$-th nearest neighbor plot ($k$-dist plot), then we can use $k+1$ as our $MinPts$ and the distance at the boundary (i.e., the elbow point) as our $\epsilon$. This method was suggested by \cite{ester1996kdd} and used by \cite{shou2019dbscan} to get the members of the Hyades open cluster.

One issue with $k$-dist plot is that it requires a visual inspection of the plot and thus can be subjective. Especially when we have a large dataset (like Gaia) and a larger fraction of field stars, it becomes more uncertain to pinpoint an exact elbow point from the plot. Due to this reason, we did not  use  the $k$-dist plot as our final metric to choose optimal parameters. However, we used $k$-dist plot to have an idea of possible ranges of $MinPts$ and $\epsilon$, outside of which the $k$-dist plots clearly do not show any boundary. Then we run the DBSCAN model for all combinations of $MinPts$ and $\epsilon$ in this range and measure their performance using $MNN$ and $MSS$ metrics. 


\subsubsection{Mean Nearest Neighbour ($MNN$) Distance}

$MNN$ distance is defined as the mean distance of the nearest neighbor of each star in the group. As the clusters are compact and members are close to each other in the parameter space, their $MNN$ distance would be small for clusters.
We finalized the value of $\epsilon$ and $Minpts$ using the elbow plots as shown for NGC~6005 in Fig.\ref{mnn}, where we get low values for $MNN$ distance. We also determined the number of retrieved members for different pairs of $\epsilon$ and $Minpts$. When the $MNN$ distance for two or more pairs is very similar, we chose the one with the highest number of members. For all twelve clusters, we found different optimal values of $\epsilon$ and \textit{Minpts} as reported in Table \ref{dbpara}. This was used in the analysis.

\subsection{Modified Silhouette Score (MSS)}

The Silhouette Score is a metric used to calculate the performance of a given clustering technique and validate the clustering algorithm. When using the silhouette approach, each point's silhouette coefficients are calculated, which indicates how much a point resembles its own cluster in relation to other clusters by offering a concise graphic illustration of how accurately each cluster has been identified.

The silhouette score is a representation of how far away an object is from other clusters (separation) compared to its own cluster (cohesion). A high value means the object is well-matched to its own cluster and poorly matched to neighboring clusters. The value of the silhouette runs between $[1, -1]$. The clustering process is useful if the majority of the items have high values. A clustering setup may have too many or too few clusters if a large number of points have low or negative values. The silhouette coefficient for the sample point is defined as,

$$s = \frac{b-a}{\max(a,b)}$$
\noindent
where $a$ is the mean distance between a sample and all other points in the same class and $b$ is the mean distance between the sample and all other points in the next class.

The problem in this metric is that the silhouette score assumes that all clusters are dense and well separated, which is not true in the case of star clusters. We have a member set, that is compact and dense in all the feature-variable spaces. But our field stars are random and uniform in all feature spaces and is mixed with the member set in feature space.  
Even when we have a  good separation of member and field stars, for a member star the value of $b$ would be close to the value of $a$, as the field stars are uniformly found all around the member set. Hence, the silhouette score will be close to 0. 

We shall use another property of our sample in this case. The standard deviation ($\sigma$) of the members would be small due to the cluster's compact nature. On the other hand, the field stars are dispersed evenly, which means that their $\sigma$ ought to be high. Therefore, in our case, $\sigma$ as a metric may be more helpful than the comparison of the distance between clusters. Therefore, a new metric is proposed for evaluating performance for clustering by an unsupervised model by making certain adjustments to the silhouette score. This newly proposed metric is named as the Modified Silhouette Score (MSS) and is denoted by

\begin{equation}
    MSS = \frac{1}{k} \sum_{i=1}^{k} \frac{(\sigma_{i, field} - \sigma_{i, member})}{\max(\sigma_{i, field}, \sigma_{i, member})}
    \label{eq:mss}
\end{equation}
\noindent
where $k$ is the total number of features and  $\sigma_{i, field}$ and $\sigma_{i, member}$ denote the $\sigma$ of the feature $i$ for field stars and members  respectively \citep{mahmudunnobe2024using}.

We would expect that a well-performed model would show the members to be distributed normally with a very small $\sigma$ and the field stars to be uniformly distributed, i.e., with a high $\sigma$.

In this case, we would have $\sigma_{field} >> \sigma_{member}$, therefore the numerator will be $\sigma_{ field} - \sigma_{i, member} \approx \sigma_{i, field}$. This will result in an MSS value very close to 1. On the other hand, for a poor performance model, the member and field stars both will have a similar random distributions. Thus, the numerator will be close to 0, resulting in an MSS value around 0. One special case is when the predicted field star group shows a stronger normal distribution (thus having low $\sigma$), but the predicted member group is distributed randomly (a larger $\sigma$). In this case, the numerator will be $\sigma_{field} - \sigma_{i, member} \approx -\sigma_{i, member}$ and the MSS value will be around -1. So, a strong negative MSS value will likely indicate that the model was able to distinguish well between member and field stars, but it mislabeled the groups. The predicted member group is the field star group and vice-versa.

\begin{figure}[h]
     \begin{center}
\includegraphics[width=0.5\textwidth,height=0.4\linewidth]{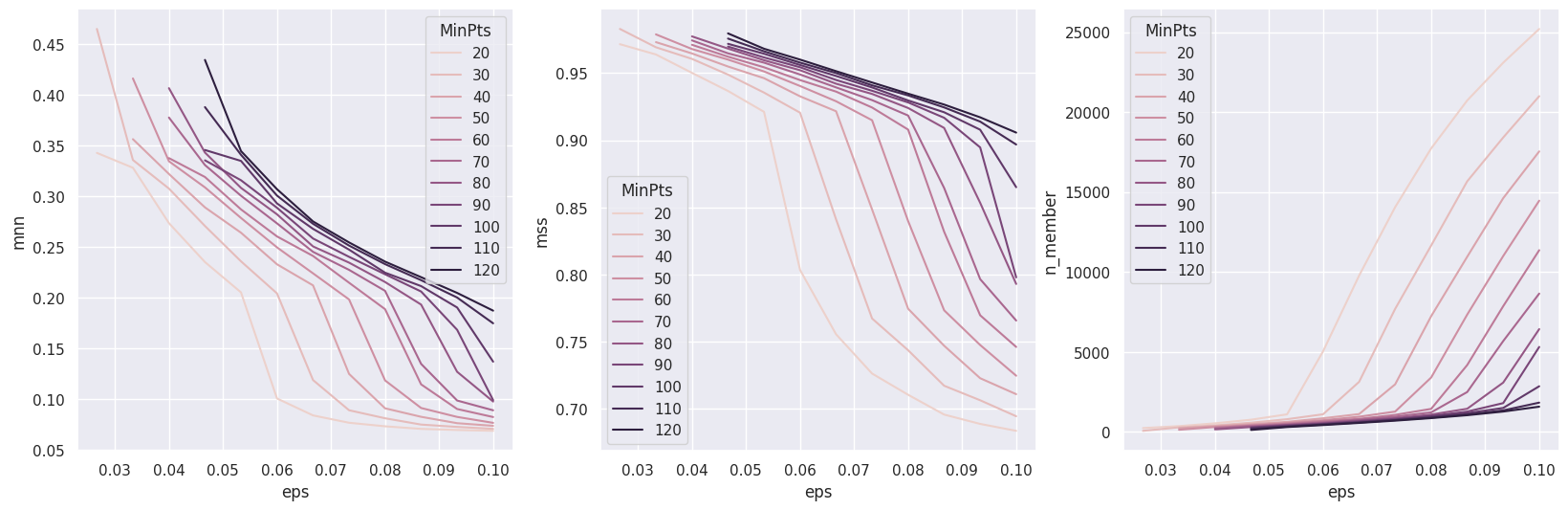}
\caption[Parameters $\epsilon$ and $Minpts$]{Parameters $\epsilon$ and $Minpts$ optimization using the $MNN$ distance and no of members for NGC~6005.}
 \label{mnn}
 \end{center}
 \end{figure}

Figure \ref{mnn} shows the plots of $\epsilon$ and $Minpts$  for NGC~6005. Here, we choose $\epsilon$ (= 0.06) that gives a small  $MNN$ distance as this is a cluster. We also note that with $\epsilon$ (= 0.06), our $MSS$ value exceeds 0.9. Further in the third plot, we notice the elbow first appearing fon $MinPts=30-40$. Hence we take $MinPts=35$.  

\section{DBSCAN Results}
\label{sec4}
Table \ref{dbpara} describes the values of $\epsilon$ and \textit{Minpts} selected  and the parallax obtained for our sample of clusters. 

\begin{table}[h]
\small
\caption{DBSCAN Parameters of the sample data}
 \label{dbpara}
\begin{tabular}{llcccc}
\hline\\
Cluster&   RA & DEC & Epsilon & $MinPts$ & Parallax \\
        & deg& deg& $\epsilon$ &   &  mas  \\
\hline
\hline
\vspace{0.5cm}
NGC 2264 &  100.2  &  9.8  &    0.06   & 20  & 1.38 $\pm$0.08\\
\vspace{0.5cm}
NGC 2682  & 132.8 & 11.8  & 0.06 & 35 &  1.15 $\pm$0.07 \\ 
\vspace{0.5cm}
NGC 2244  &  98.0  &  4.9  &  0.065 &  25  &  0.66 $\pm$0.09  \\
\vspace{0.5cm}
NGC 3293  & 80.7  &  33.4  &  0.065  &  25  &  0.39  $\pm$0.05 \\
\vspace{0.5cm}
NGC 6913  &  305.9  & 38.5  &  0.05  &  40 &  0.56 $\pm$0.05 \\
\vspace{0.5cm}
NGC 7142  &  326.3  & 65.8  &  0.06  &  20  &  0.40 $\pm$0.06 \\
\vspace{0.5cm}
IC 1805 &  38.2  &  61.5  &  0.05  &  30  &  0.41 $\pm$0.06 \\
\vspace{0.5cm}
NGC 6231  &  235.5 & -41.8 & 0.04  &  20 &  0.60 $\pm$0.07\\
\vspace{0.5cm}
NGC 2243   &  97.4  &  -31.3 &  0.06  &  15 &  0.25 $\pm$0.05\\
\vspace{0.5cm}
NGC 6451 & 267.675 & -30.206  & 0.07 & 100 & 0.35 $\pm$0.06\\
\vspace{0.5cm}
NGC 6005 & 238.955 & -57.439 & 0.07 & 100 & 0.38 $\pm$0.05\\ 
\vspace{0.5cm}
NGC 6583 & 273.962 & -22.143 & 0.06 & 50 & 0.41 $\pm$0.05 \\

\hline
\end{tabular}
\end{table}

The results for memberships obtained after using DBSCAN algorithm are shown in Table \ref{dbresults} and compared with the CG members.


\begin{table}[h]
\small
 \caption{Performance of DBSCAN for sample clusters}
    \centering
\begin{tabular}{lrrrrrr}
\hline
  Cluster & $MNN$ & $MSS$ &  Member &  Member & Ratio & Sample\\
    &   & & D & CG & D/CG & after filters\\
\hline
\hline
\vspace{0.5cm}
NGC 2264  & 0.33 & 0.94 &    325 &   186 & 1.75 & 4829\\
  \vspace{0.5cm}
NGC 2682  & 0.23 & 0.95 &  1222 &  848& 1.44 & 2576 \\
  \vspace{0.5cm}
NGC 2244  & 0.19 & 0.91 &    1803 &   1701 & 1.05& 6530\\
 \vspace{0.5cm}
 NGC 3293  & 0.2 & 0.94 &   1027 &      657 & 2.23 & 27552\\
 \vspace{0.5cm}
 NGC 6913  & 0.25 & 0.95 &   824 &       170 & 4.85 & 13660\\
 \vspace{0.5cm}
 NGC 7142  & 0.23 & 0.93 &   813 &      539 & 1.50 & 9820\\
 \vspace{0.5cm}
  IC 1805  &  0.18 & 0.92 &    1252 &       456& 2.74 & 9472\\
  \vspace{0.5cm}
 NGC  6231 & 0.17 & 0.92 &  2032 &   1560 &  1.29 & 24895\\
 \vspace{0.5cm}
 NGC 2243  & 0.29 & 0.96 &  548 &       531 & 1.03 & 5380 \\
\vspace{0.5cm}
NGC 6451 &  0.20 & 0.93 & 1560 & 1171 & 1.33& 50422 \\
\vspace{0.5cm}
NGC 6005 &  0.21 & 0.92 & 1072 & 708 & 1.5&  50570\\ 
\vspace{0.5cm}
NGC 6583 & 0.3 & 0.94 & 455 & 364 & 1.25 & 47821 \\
\hline

\end{tabular}

    \label{dbresults}
\end{table}

\begin{figure}[h]
   \centering
    \includegraphics[width = 
    \linewidth]{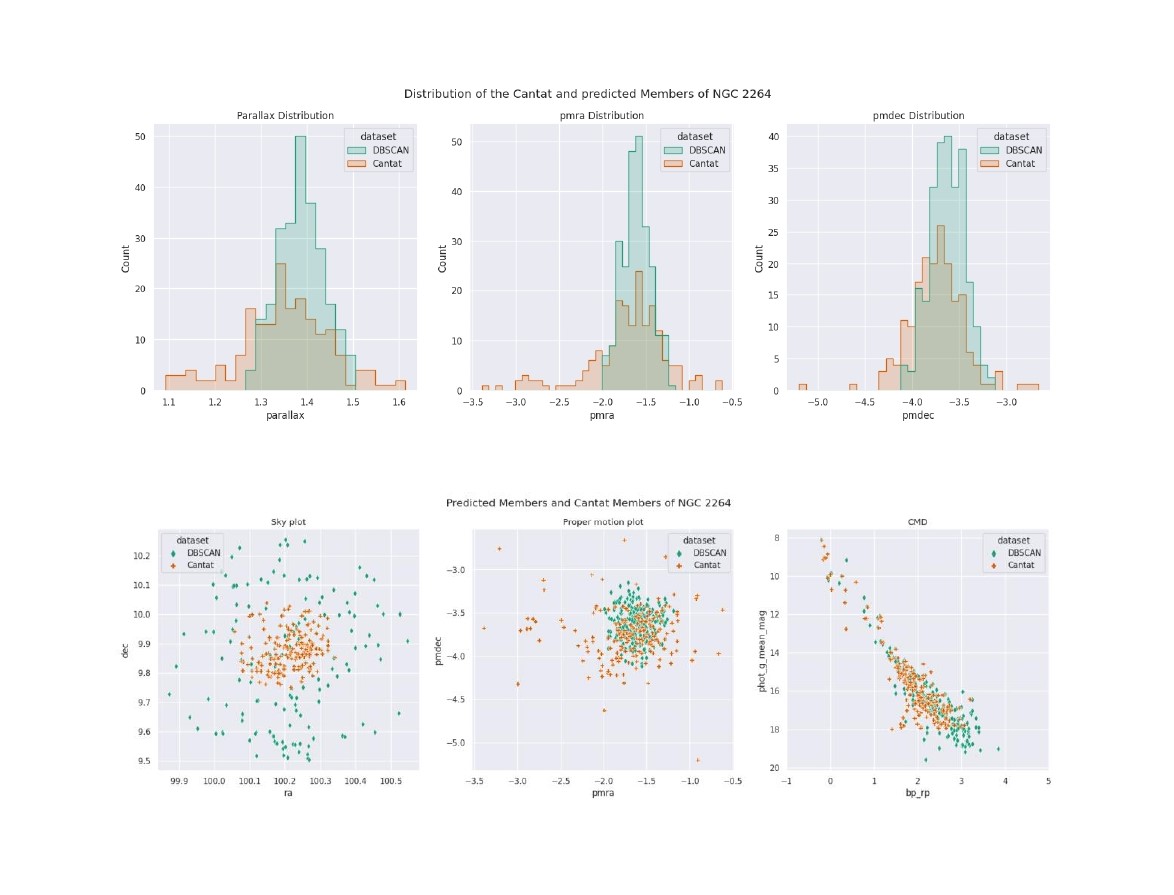}
    \caption[Revised members of NGC~2264]{Revised members of NGC~2264 CG(orange) and DBSCAN (green).}
    \label{d2264}
\end{figure}

\begin{figure}[h]
   \centering
    \includegraphics[width = 
    \linewidth]{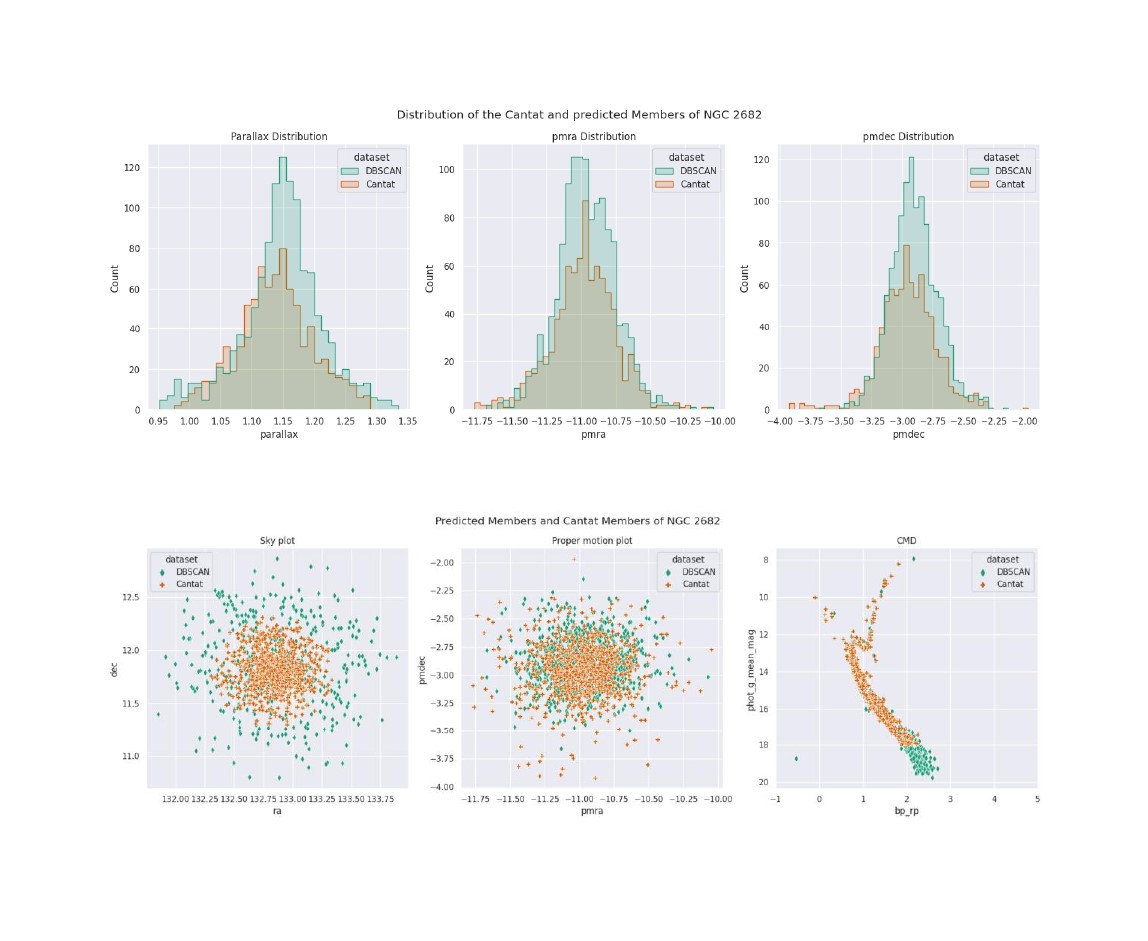}
    \caption[Revised members of NGC~2682]{Revised members of NGC~2682 CG(orange) and DBSCAN (green).}
    \label{d2682}
\end{figure}

\begin{figure}[h]
   \centering
    \includegraphics[width = 
    \linewidth]{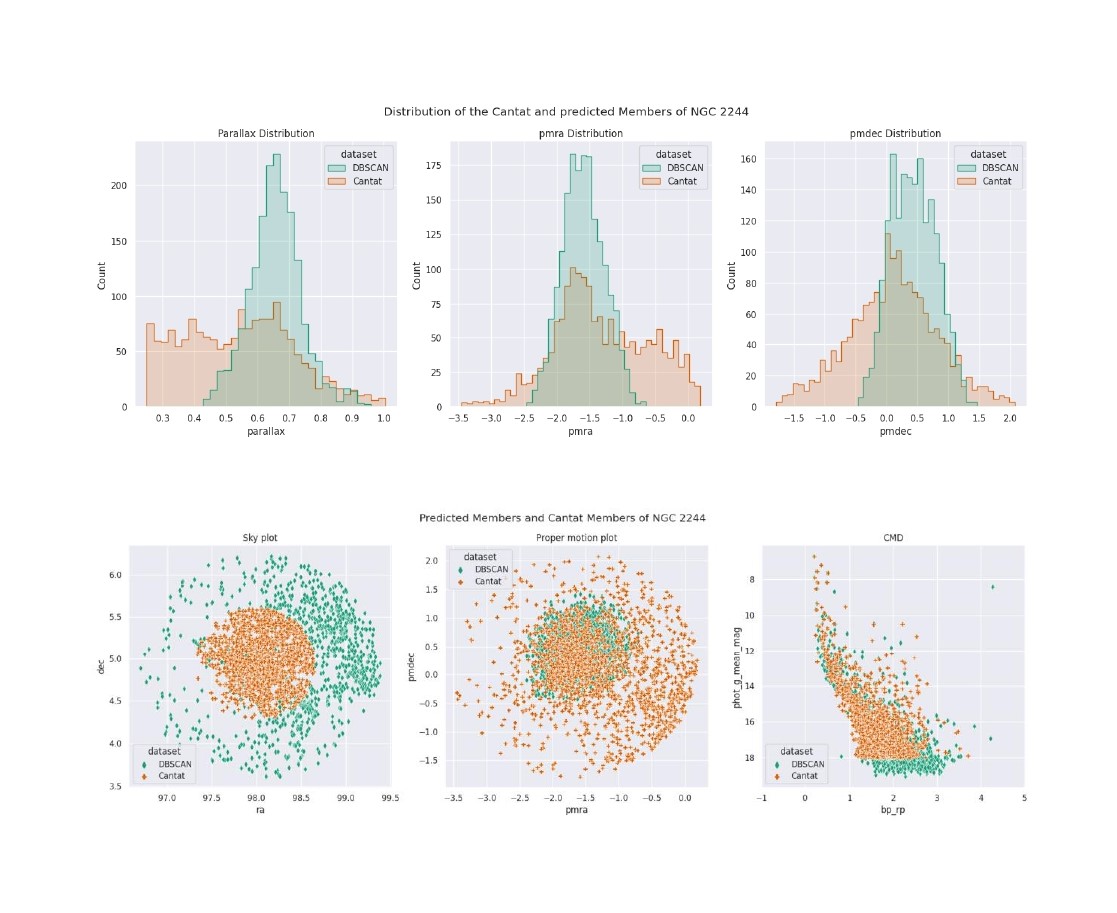}
    \caption[Revised members of NGC~2244]{Revised members of NGC~2244 CG(orange) and DBSCAN (green).}
    \label{d2244}
\end{figure}

\begin{figure}[h]
   \centering
    \includegraphics[width = 
    \linewidth]{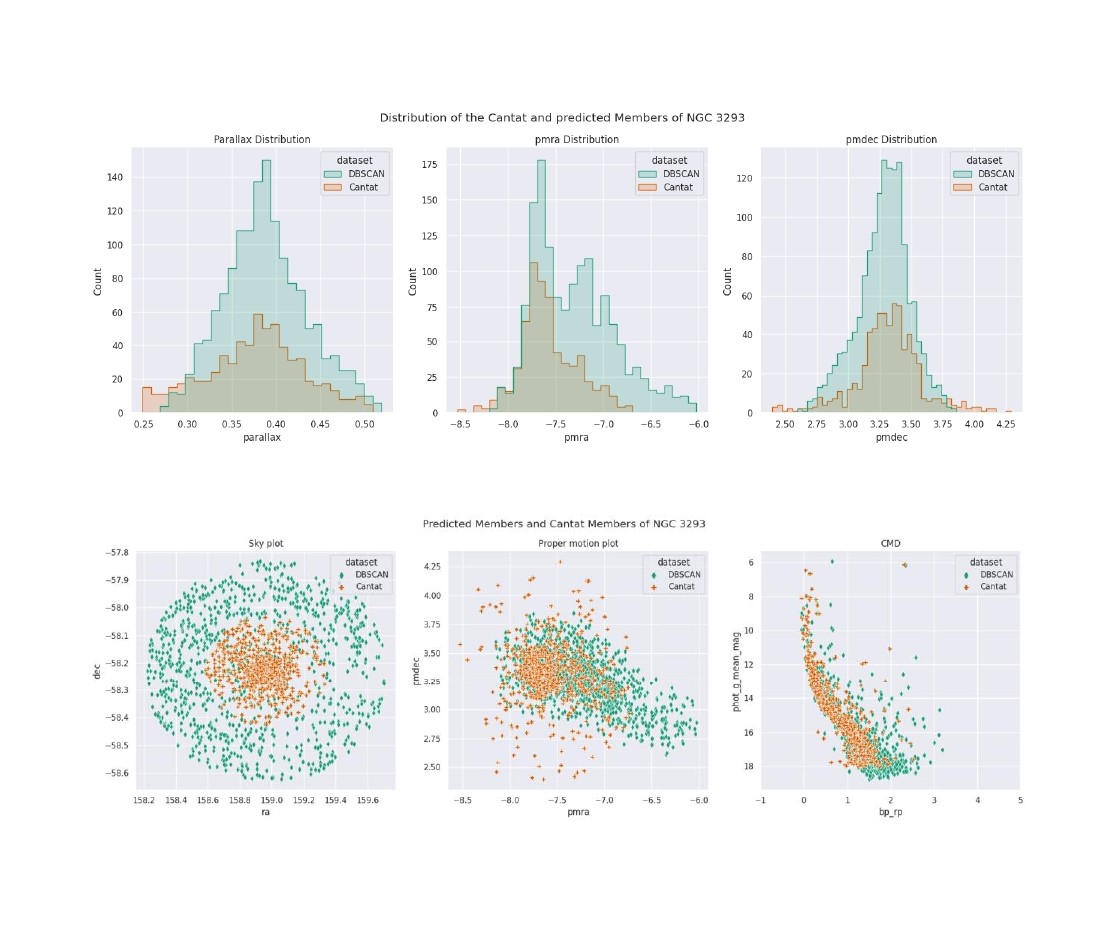}
    \caption[Revised members of NGC~3293]{Revised members of NGC~3293 CG(orange) and DBSCAN (green). }
    \label{d3293}
\end{figure}

\begin{figure}[h]
   \centering
    \includegraphics[width = 
    \linewidth]{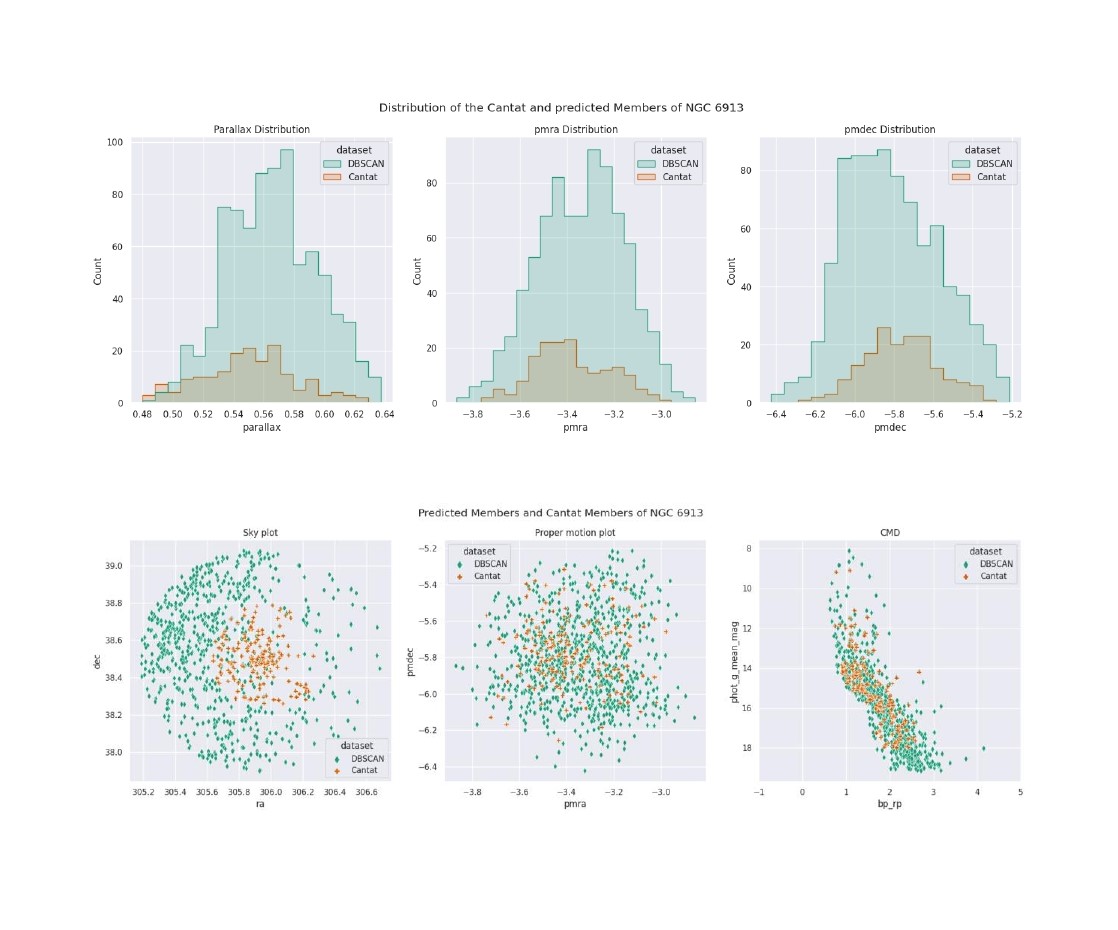}
    \caption[Revised members of NGC~6913]{Revised members of NGC~6913 CG(orange) and DBSCAN (green). } 
    \label{d6913}
\end{figure}

\begin{figure}[h]
   \centering
    \includegraphics[width = 
    \linewidth]{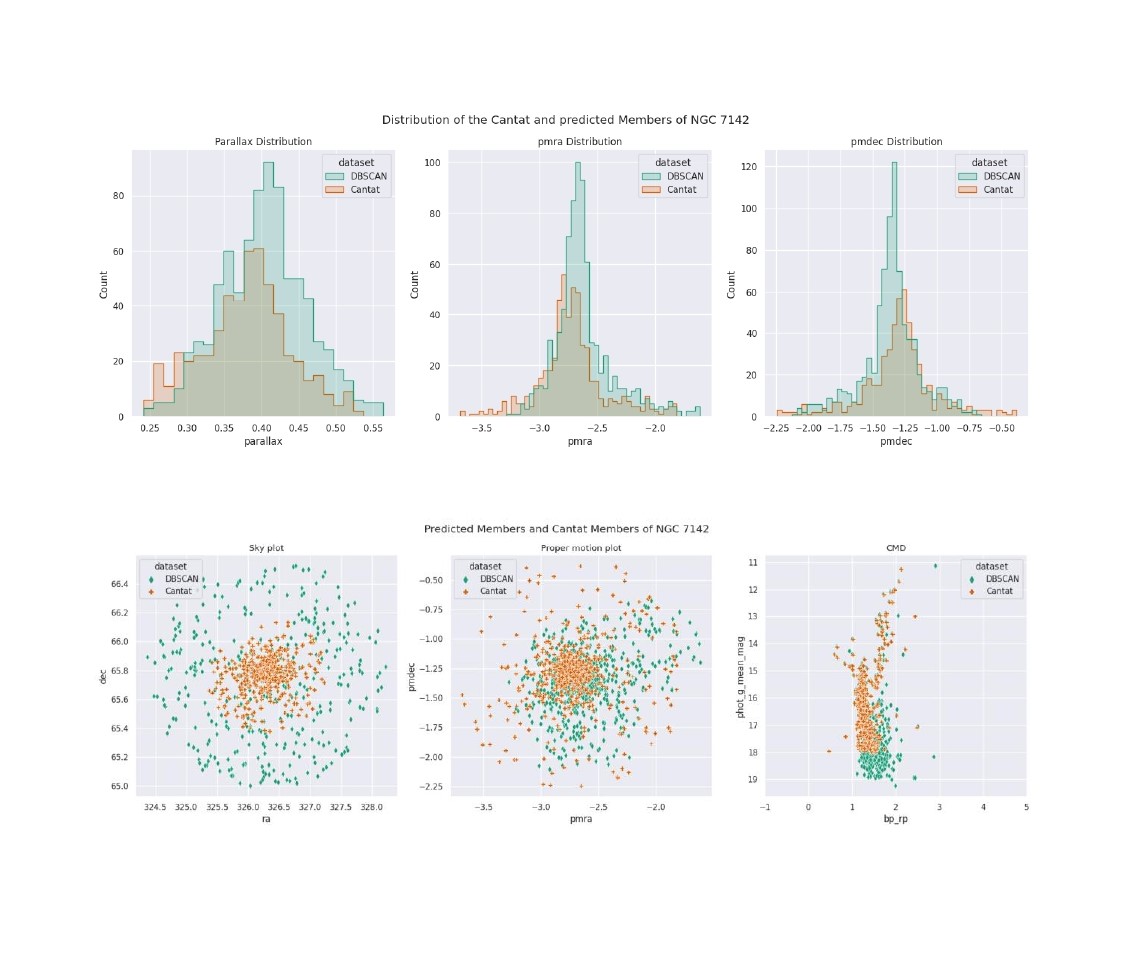}
    \caption[Revised members of NGC~7142]{Revised members of NGC~7142 CG(orange) and DBSCAN (green). }    \label{d7142}
\end{figure}

\begin{figure}[h]
   \centering
    \includegraphics[width = 
   \linewidth]{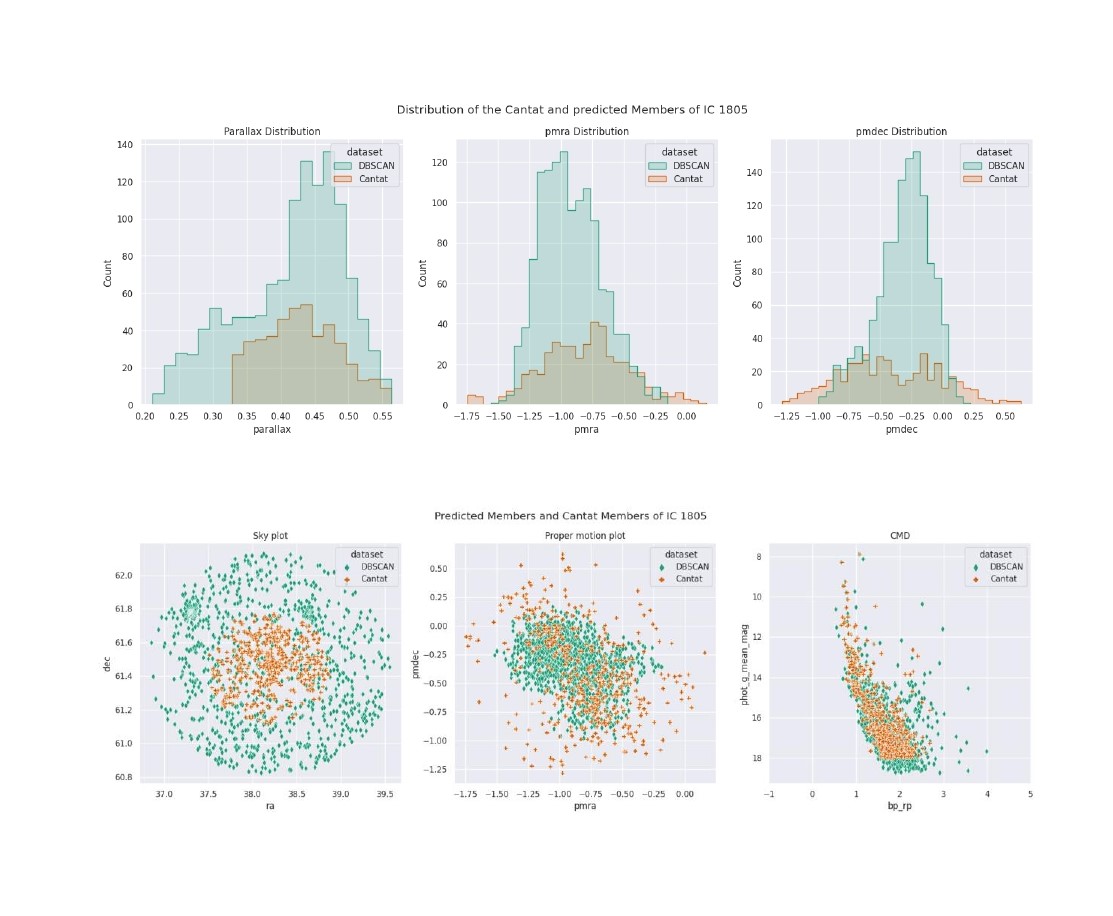}
    \caption[Revised members of IC~1805]{Revised members of IC~1805 CG(orange) and DBSCAN (green).} 
    \label{d1805}
\end{figure}

\begin{figure}[h]
   \centering
    \includegraphics[width = \linewidth]{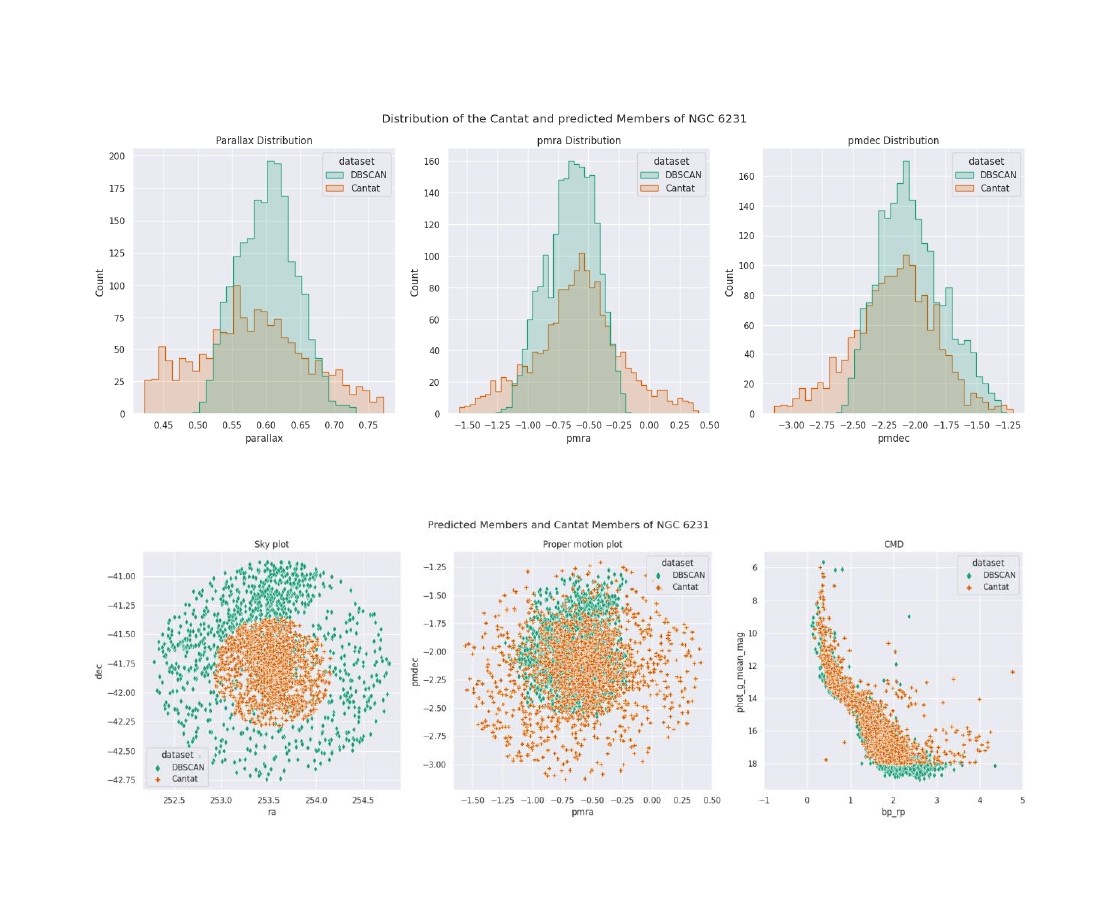}
    \caption[Revised members of NGC~6231]{Revised members of NGC~6231CG(orange) and DBSCAN (green). }  
    \label{d6231}
\end{figure}

\begin{figure}[h]
   \centering
    \includegraphics[width =\columnwidth]{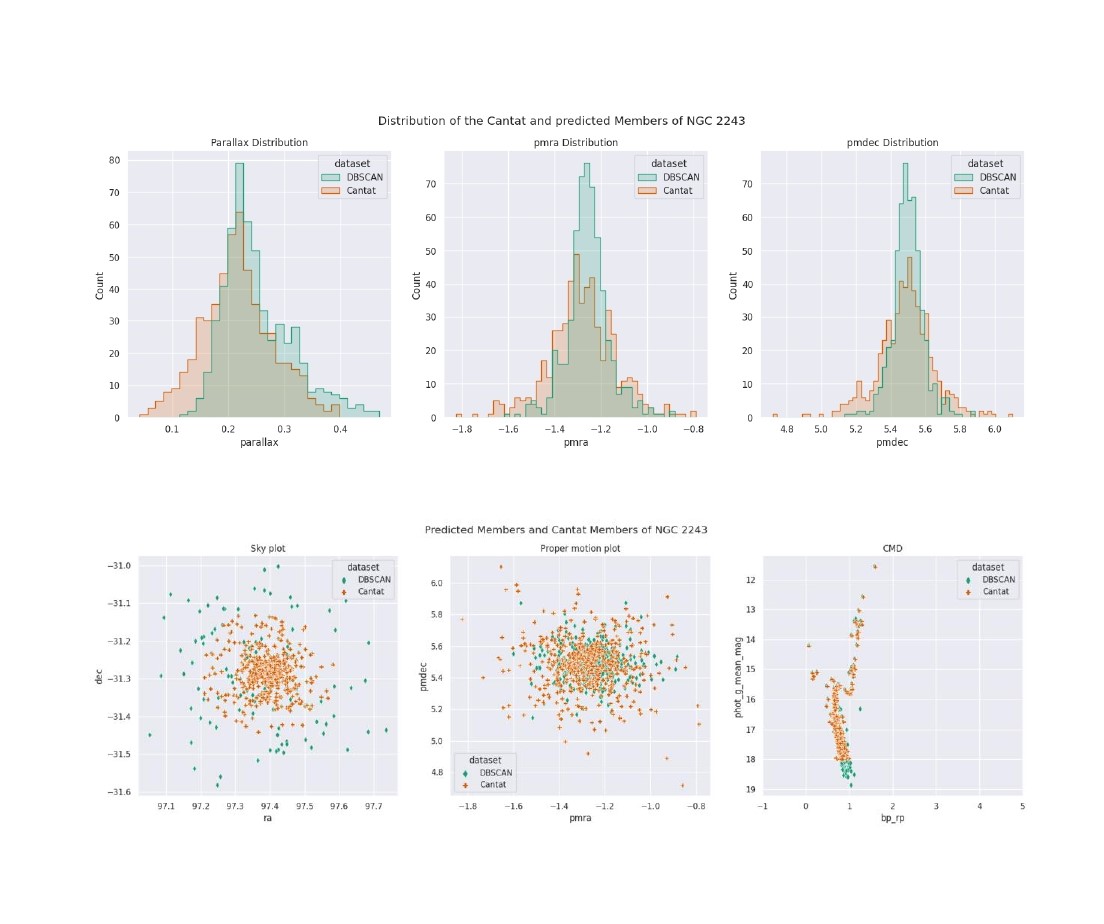}
    \caption[Revised members of NGC~2243]{Revised members of NGC~2243 CG(orange) and DBSCAN (green).}
    \label{d2243}
\end{figure}

\begin{figure}[h]
   \centering
    \includegraphics[width =\columnwidth]{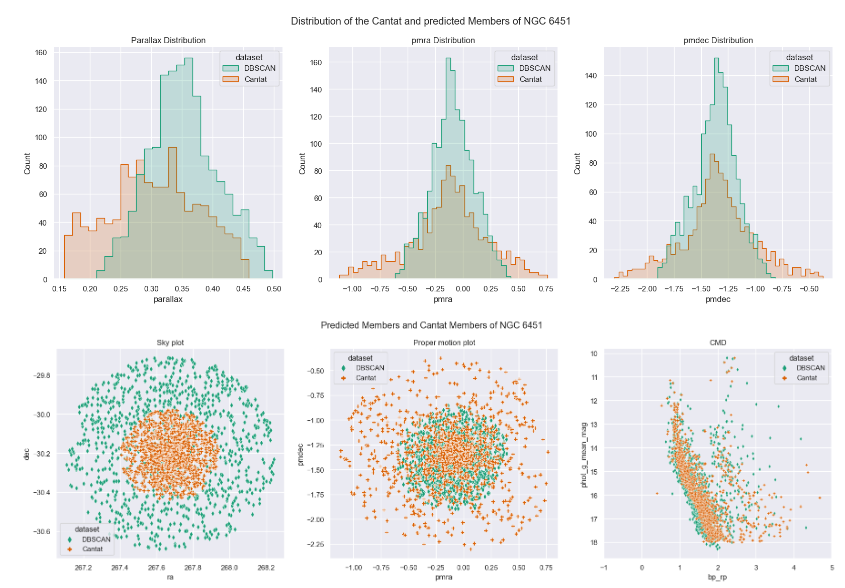}
    \caption[Revised members of NGC~6451]{Revised members of NGC~6451 CG(orange) and DBSCAN (green).}
    \label{d6451}
\end{figure}

\begin{figure}[h]
   \centering
    \includegraphics[width =\columnwidth]{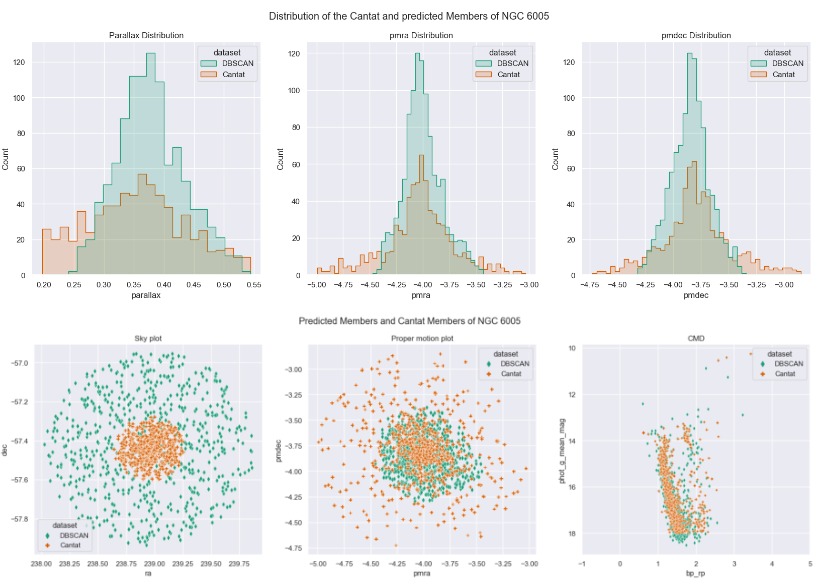}
    \caption[Revised members of NGC~6005]{Revised members of NGC~6005 CG(orange) and DBSCAN (green).}
    \label{d6005}
\end{figure}

\begin{figure}[h]
   \centering
    \includegraphics[width =\columnwidth]{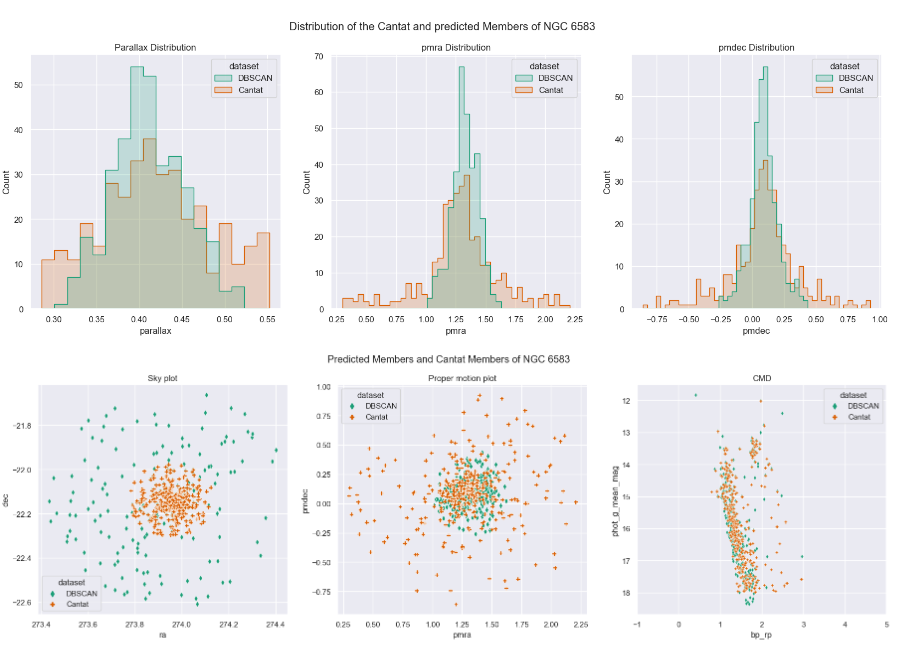}
    \caption[Revised members of NGC~6853]{Revised members of NGC~6853 CG(orange) and DBSCAN (green).}
    \label{d6853}
\end{figure}

Figures \ref{d2264} to \ref{d6853} show plots of the revised membership in our sample of clusters where CG(orange) and DBSCAN (green). The upper left plot shows the histogram of the parallax distribution of the CG members and the members found by our sample. The upper middle plot and the upper right plots show the histogram of the  proper motion distributions of $pmra$ and $pmdec$. Our sample has a narrower distribution of parallax and proper motions with a smaller standard deviation as we have used DR3 data, while CG is based on DR2 data. The lower plots show the sky plot (lower left), the proper motion plot (lower middle) and the color-magnitude diagram (lower right) of the CG(orange) and DBSCAN (green) members. The plot shows we have explored the outer regions of the clusters and have found fainter members down to $20^m$. 
\cite{10.1093/mnras/stad1589} used DBSCAN and GMM to find members of 12 open clusters. Of these we have in common two clusters: NGC~2682 (M67) and NGC~2243. And our values are in good agreement as shown in Table \ref{comp}.

\begin{table}[h]
\small
 \caption{Comparison of DBSCAN results with \cite{10.1093/mnras/stad1589}(NM)}
\centering
\begin{tabular}{lllll}
\hline
  Cluster  &Parallax & pmra &  pmdec &  distance \\ 
   &mas & mas yr$^{-1}$ &  mas yr$^{-1}$ &  pc \\ 
\hline
 & & NGC~2682 &  &  \\
\hline
Ours &1.14$\pm$0.06 & −10.96$\pm$0.2 & −2.90$\pm$0.19 &  877.1 \\
\vspace{0.5cm}
NM& 1.15$\pm$0.09 & −10.96$\pm$0.09 & −2.90$\pm$0.07 &  864.3 \\
\hline
 & & NGC~2243 &  &  \\
\hline
Ours    & 0.25$\pm$0.06 & −1.25 $\pm$ 0.09 & 5.49$\pm$ 0.05 & 4000 \\
\vspace{0.5cm}
 NM     &0.23 $\pm$ 0.08 & −1.26$\pm$0.09 & 5.50$\pm$0.09 & 3660.6 \\
\hline
\end{tabular}
\label{comp}
\end{table}

In Fig. \ref{d3293} for NGC~3293, it is visible that even Cantat-Gaudin data has a hint of a double peak in pmra. In our case, this peak has got more enhanced. There is definitely some sub-structure in NGC~3293 in the velocity space that needs to be explored in detail.  

\section{Spectroscopic Data: APOGEE and GALAH}\label{sec5}
As the member stars are born from the same molecular cloud, their chemical abundances should be similar. 
We compared the chemical abundances of the DBSCAN members and the members  found by \cite{cantat2018gaia} using APOGEE and GALAH data, where available. 
Apache Point Observatory Galactic Evolution Experiment(APOGEE)
is a large scale, stellar spectroscopic survey which is conducted in the near infra-red (IR) region of the electromagnetic spectrum.
APOGEE \citep{majewski2017apache} observations provide $R \sim 22,500 $ spectra in the infrared H-band, $1.5-1.7\mu m$, as part of the third and fourth phases of the Sloan Digital Sky Survey \citep{eisenstein2011sdss, blanton2017sloan}.

\begin{figure}[h]
     \begin{center}
\includegraphics[width=1.2\linewidth,height=1.0\linewidth]{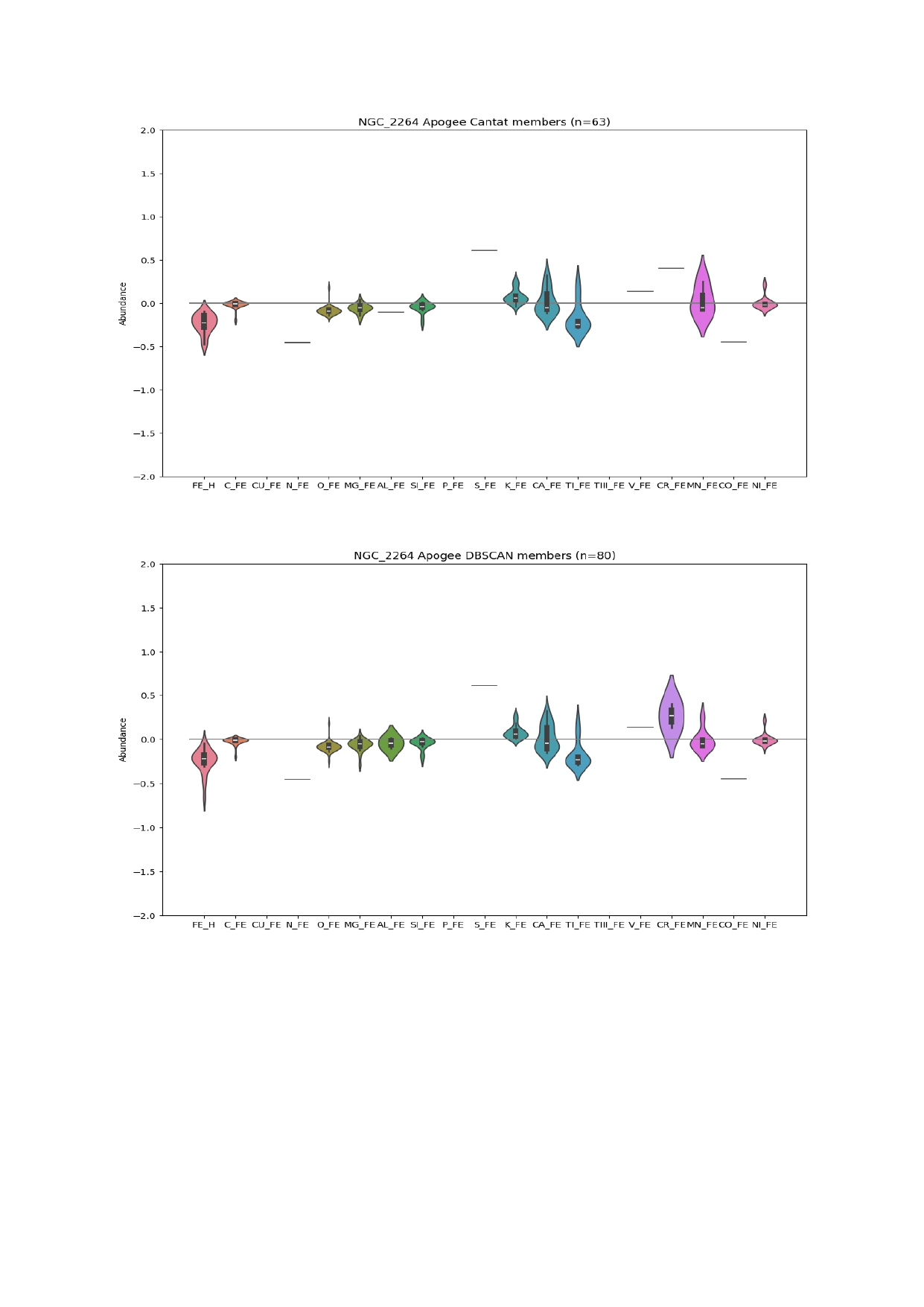}
\caption[APOGEE data for NGC~2264]{Chemical abundances of members from APOGEE for NGC2264  (a) Upper plot \citep{cantat2018gaia} (b) Our results
}
\label{a2264}
\end{center}
\end{figure}

Figures \ref{a2264}  show the chemical abundances of members  of NGC~2264 from APOGEE. The upper plots shows members found by \cite{cantat2018gaia} and the lower ones are our results. The rest of the plots are in Appendix \ref{a1}.

The Galactic Archaeology with HERMES (GALAH) is a high resolution, ground-based spectroscopic survey.
It is carried out using the Anglo-Australian Telescope's Two Degree Field (2dF) of view and the High Efficiency and Resolution Multi-Element Spectrograph (HERMES) \citep{barden2010hermes, heijmans2012integrating, sheinis2015first}.The HERMES spectrograph gives a high resolution (R $\sim$ 28000) spectra for 392 stars in four passbands.
Table \ref{APOGEE} shows the number of stars that matched with \citep{cantat2018gaia} and DBSCAN members from APOGEE and GALAH surveys.
\begin{table}[h]
\caption{APOGEE and GALAH matches with cluster sample}
\label{APOGEE}
\small
\begin{tabular}{llllc}
\hline\\
\vspace{0.5cm}

Cluster& APOGEE  & APOGEE & GALAH   & GALAH  \\
& CG& DBSCAN& CG& DBSCAN  \\
\hline\\
\vspace{0.5cm}
NGC 2264 &  71 &  203 &   0   & 0  \\
\vspace{0.5cm}
NGC 2682  & 430 & 405 & 274 & 279 \\ 
\vspace{0.5cm}
NGC 2244  &  191  &  200  &  0 & 0 \\
\vspace{0.5cm}
NGC 3293  & 25  &  69 &  24 &  88  \\
\vspace{0.5cm}
NGC 6913  &  60  & 40  &  0 &  0 \\
\vspace{0.5cm}
IC 1805 &  55  &  84  &  0 &  0 \\
\vspace{0.5cm}
NGC 2243   &  13  &  15 &  8  &  10 \\  
\vspace{0.5cm}
NGC~6451&. 4 & 6  & 0 & 0\\
\vspace{0.5cm}
NGC~6583 & 0 & 0  &  23 & 22 \\ 
\hline
\end{tabular}

\end{table}

The figures show a good agreement in chemical abundances obtained by \citep{cantat2018gaia} and DBSCAN members obtained by us. 

\begin{figure}[h]
     \begin{center}
   \includegraphics[width=1.2\linewidth,height=1.0\linewidth]{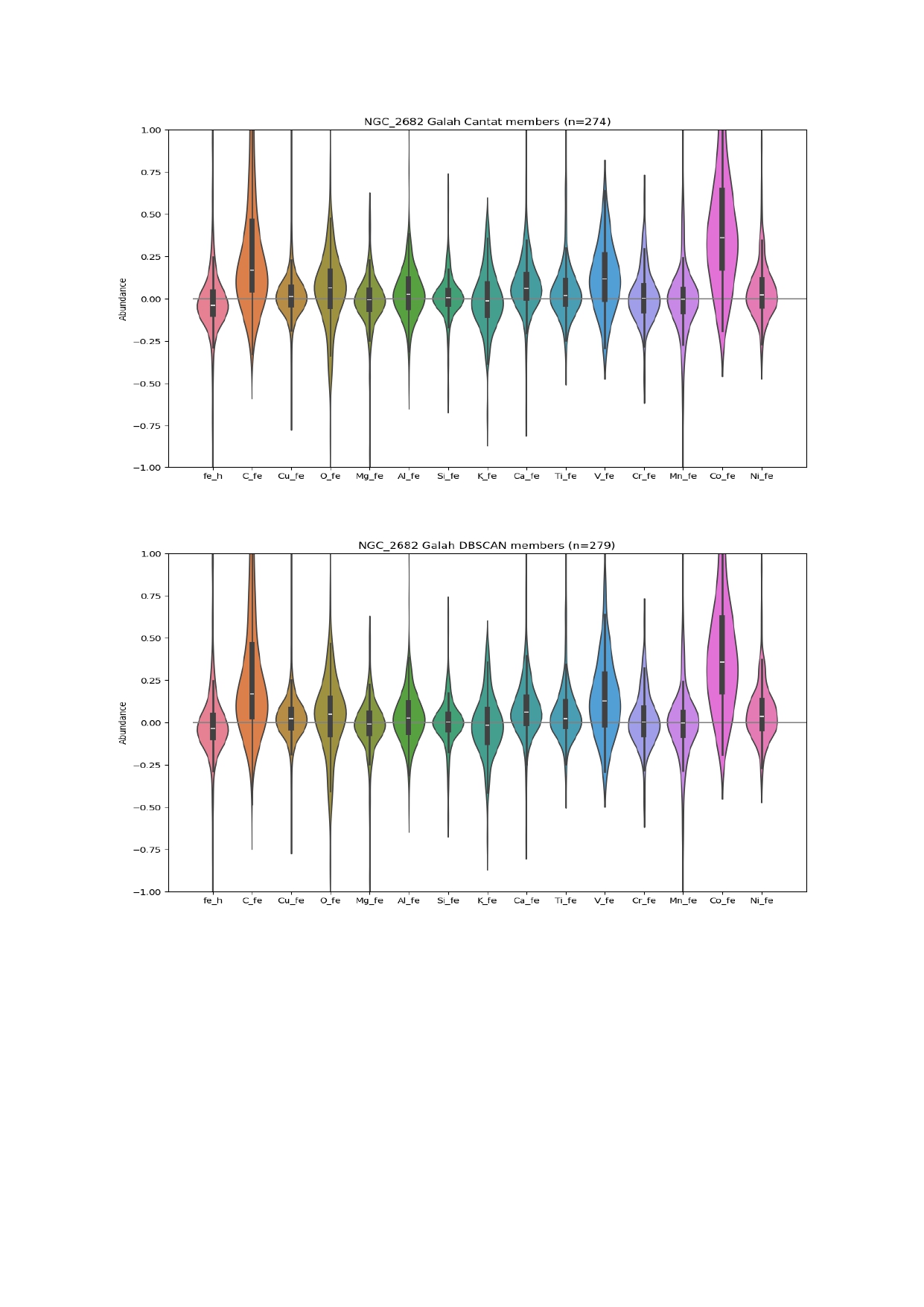}
    \caption[GALAH data for NGC~2682]{Chemical abundances of members from GALAH for NGC 2682 (a) Upper plot \citep{cantat2018gaia} (b) Our results
}
\label{G2682}
\end{center}
\end{figure}

Figures \ref{G2682} shows the chemical abundances of members of NGC~2682 from GALAH. The upper plot shows members found by \citep{cantat2018gaia} and the lower one is our result. The rest of the plots are in Appendix \ref{a1}.

\section{ASteCA Results}\label{sec6}
ASteCA \cite{perren2015asteca} is a software package that does an automated analysis of photometric data of star clusters.
Input parameter and data files are provided to run the code in a specified format that gives an output of all necessary cluster parameters and plots.  
ASteCA performs an exhaustive analysis by using both spatial and photometric data.
The structural parameters of a star cluster, such as its exact centre location and radius value, are derived from positional data, while the remaining functions require an observed magnitude and colour.

\begin{figure}[h]
     \begin{center}
   \includegraphics[width=0.5\textwidth]{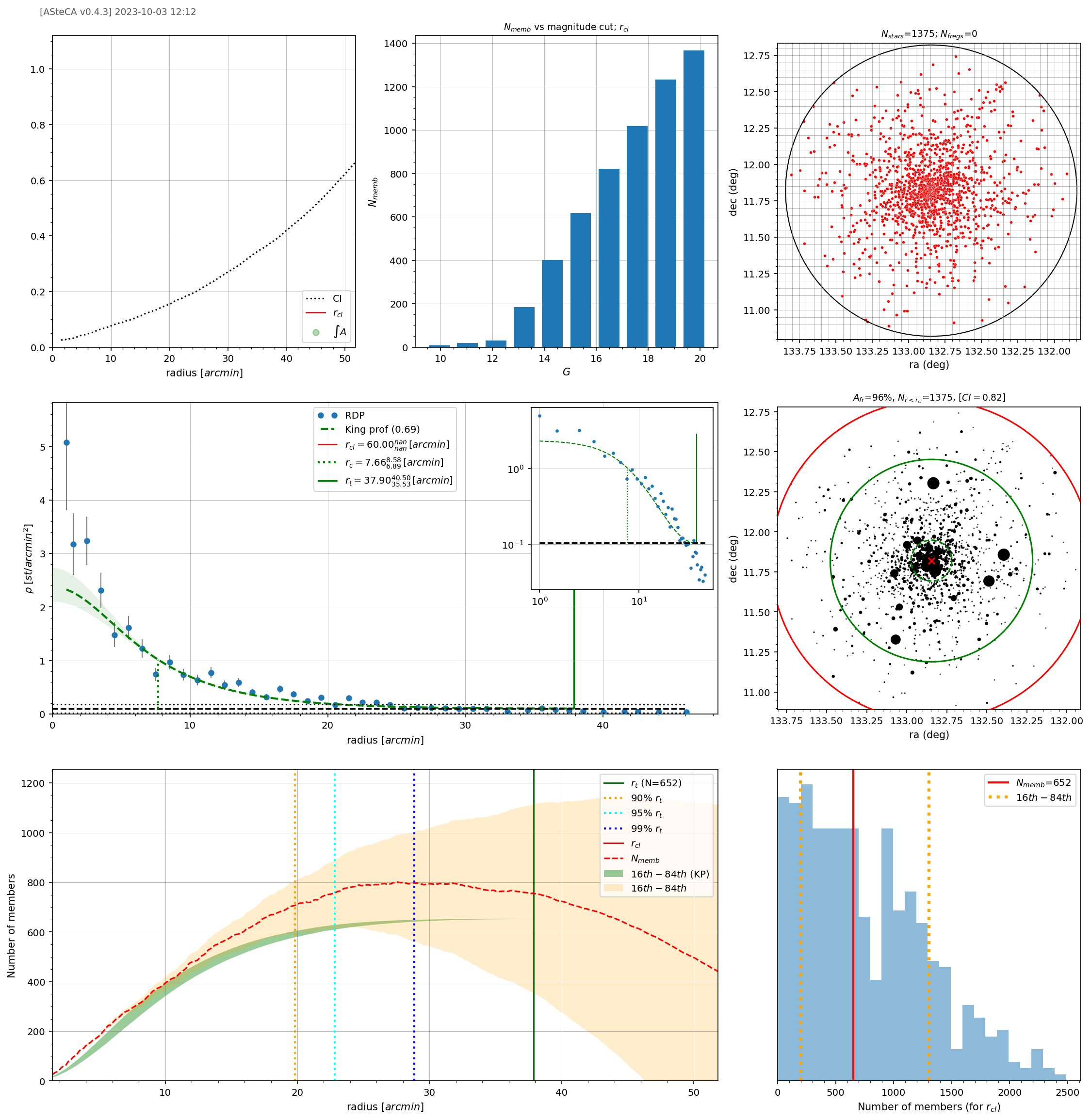}
     \caption{ASteCA plots of NGC~2682}
\label{AS2682}
\end{center}
\end{figure}

\begin{figure}[h]
     \begin{center}
   \includegraphics[width=0.5\textwidth]{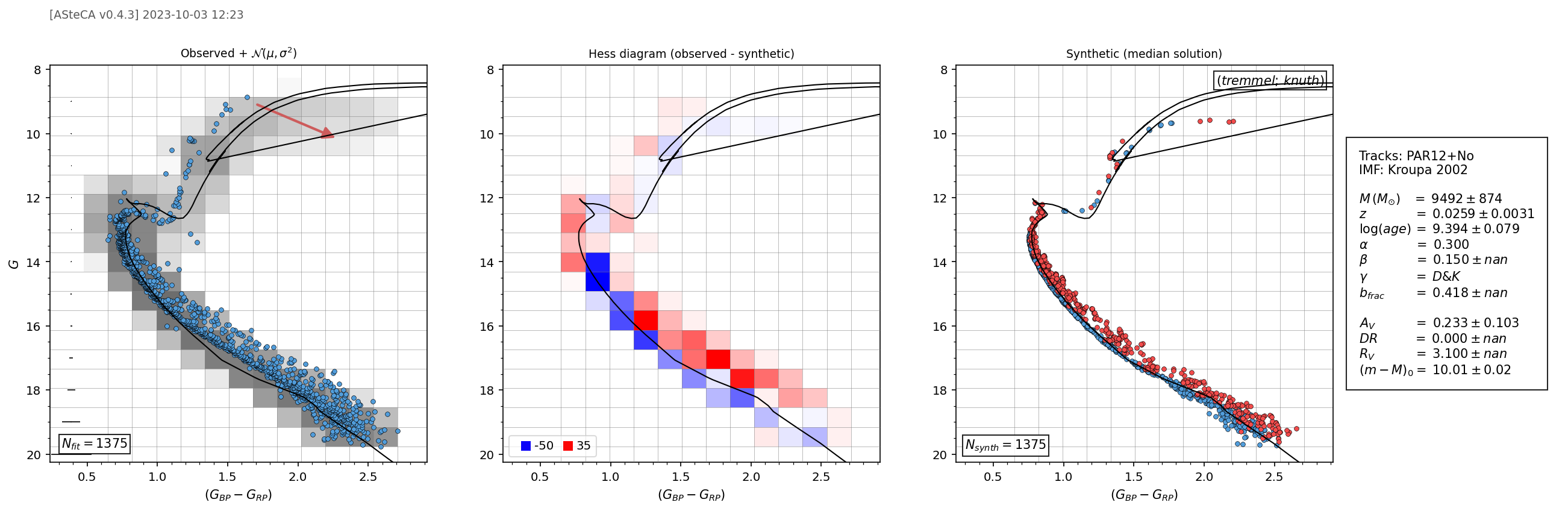}
     \caption{ASteCA CMD plots of NGC~2682}
\label{AS2682_2}
\end{center}
\end{figure}

Table \ref{Astecaparadb} shows  the ASteCA parameters of some of our sample clusters and found that they compare well with the \cite{cantat2020clusters} values.
Figures \ref{AS2682} to  \ref{AS6583_2} show  ASteCA plots  and CMDs of the revised members that we obtained after running DBSCAN clustering algorithm on our sample.

\begin{table}[h]
\begin{center}   
\small

\begin{tabular}{lcccc}
\hline
&     ASteCA & &   Cantat & \\
 Cluster &  log(age) & d &  log(age) & d\\
\hline
\hline
\\
\vspace{0.5cm}
NGC~2264 &  7.176  & 714  & 7.44 & 707 \\
\vspace{0.5cm}
NGC~2682 &  9.39 & 1000 &  9.63 & 889\\
\vspace{0.5cm}
NGC~2244 & 7.34  & 1061 &  7.1 & 1478\\
\vspace{0.5cm}
NGC~3293 &  7.4 & 1459  & 7.01 & 2710\\
\vspace{0.5cm}
NGC~6913 &  7.8 & 1318 & 7.34 & 1608\\
\vspace{0.5cm}
NGC~7142  & 9.6 & 2238 &  9.49 & 2406\\
\vspace{0.5cm}
IC~1805 &  7.547 & 2500  & 6.88 & 1964\\
\vspace{0.5cm}
NGC~6231 & 7.45 & 1032  & 7.14 & 1475\\
\vspace{0.5cm}
NGC~2243 & 9.89 & 3311  &  9.64 & 3719\\
\vspace{0.5cm}
NGC 6451 & 7.5 &  2857 & 7.41 &2777.0 \\
\vspace{0.5cm}
NGC 6005 & 9.2& 2631&9.1 &  2383.0  \\ 
\vspace{0.5cm}
NGC 6583 & 9.0 & 2439 &  9.08 &  2053.0  \\

\hline
 \end{tabular}
\caption[ASteCA parameters]{ASteCA Parameters vs parameters from \citep{cantat2020clusters} using DBSCAN}.
\label{Astecaparadb}
\end{center}
\end{table}

We have used ASteCA to find parameters of the clusters studied using the revised membership data. Figures. \ref{AS2682} - \ref{AS2682_2} show plots for NGC~2682. The rest of the plots are in the Appendix \ref{a2}. 



\section{Conclusion and Discussion}
\label{sec7}
DBSCAN is an unsupervised method used to identify clusters in the data. 
In this work, we used the DBSCAN algorithm to find the membership of stars in twelve open clusters (NGC~2264, NGC~2682, NGC~2244, NGC~3293, NGC~6913, NGC~7142, IC~1805, NGC~6231, NGC~2243, NGC 6451, 
NGC 6005 and NGC 6583) based on Gaia DR3 Data.   
Since DBSCAN requires no prior knowledge about the clusters \citep{ester1996kdd}, it provides an unbiased method to identify the cluster members, compared to our earlier work using the supervised method of Random Forest \citep{mahmudunnobe2021membership}.
We made use of the elbow method using $k$-dist graphs as suggested by \citep{shou2019dbscan}, and the $MNN$ and $MSS$ metrics to find the optimal range of values of $\epsilon$ and $MinPts$. The revised list of members obtained by us increased the number upto 4.85 times and found fainter members in the outer regions of clusters. \\
We also compared the chemical abundances of our obtained members from APOGEE and GALAH and found an agreement  with the chemical abundances of the members obtained by \citep{cantat2018gaia}. Table \ref{dbresults} shows the values of $MNN$, $MSS$ and CG members and those we obtained. We also compare the values for cluster parameters obtained by us using ASteCA with the values obtained by obtained by \citep{cantat2018gaia}. We find our method to be very effective even for inner disk clusters with high field star contamination as well as for clusters of varying ages, distances and locations in the Galaxy. 

This revised sample of cluster members can be used to find the Initial Mass Function (IMF), mass segregation,  variable stars and to study features of the color magnitude diagram like gaps \cite{hasan2023gaps} and many more detailed studies of star clusters. We plan to explore these in our future work. 

\section{Acknowledgements}
This work has made use of data from the European Space Agency (ESA) mission
{\it Gaia} (\url{https://www.cosmos.esa.int/gaia}), processed by the {\it Gaia}
Data Processing and Analysis Consortium (DPAC,
\url{https://www.cosmos.esa.int/web/gaia/dpac/consortium}). Funding for the DPAC
 has been provided by national institutions, in particular the institutions
 participating in the {\it Gaia} Multilateral Agreement.
 
 We would also like to thank our anonymous referee whose valuable comments helped imrove the quality of this paper.

\bibliographystyle{elsarticle-harv} 
\bibliography{example}
\appendix

\section{Spectroscopic Data: APOGEE and GALAH data}\label{a1}

\begin{figure}[h]
     \begin{center}
   \includegraphics[width=1.2\linewidth,height=1.0\linewidth]{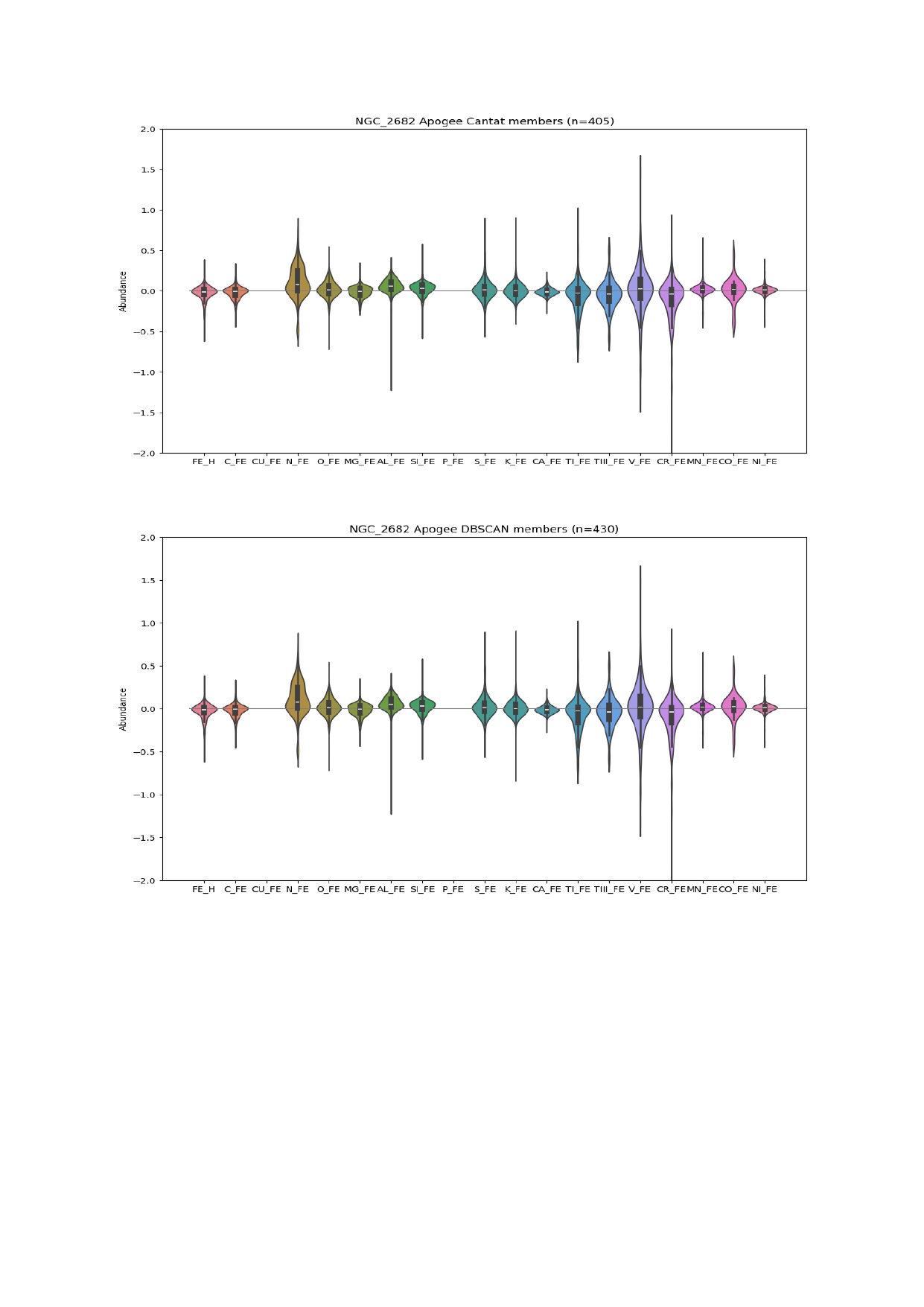}
     \caption[APOGEE data for NGC~2682]{Chemical abundances of members from APOGEE for NGC2682  (a) Upper plot \citep{cantat2018gaia} (b) Our results
}
\label{a2682}
\end{center}
\end{figure}

\begin{figure}[h]
     \begin{center}
   \includegraphics[width=1.2\linewidth,height=1.0\linewidth]{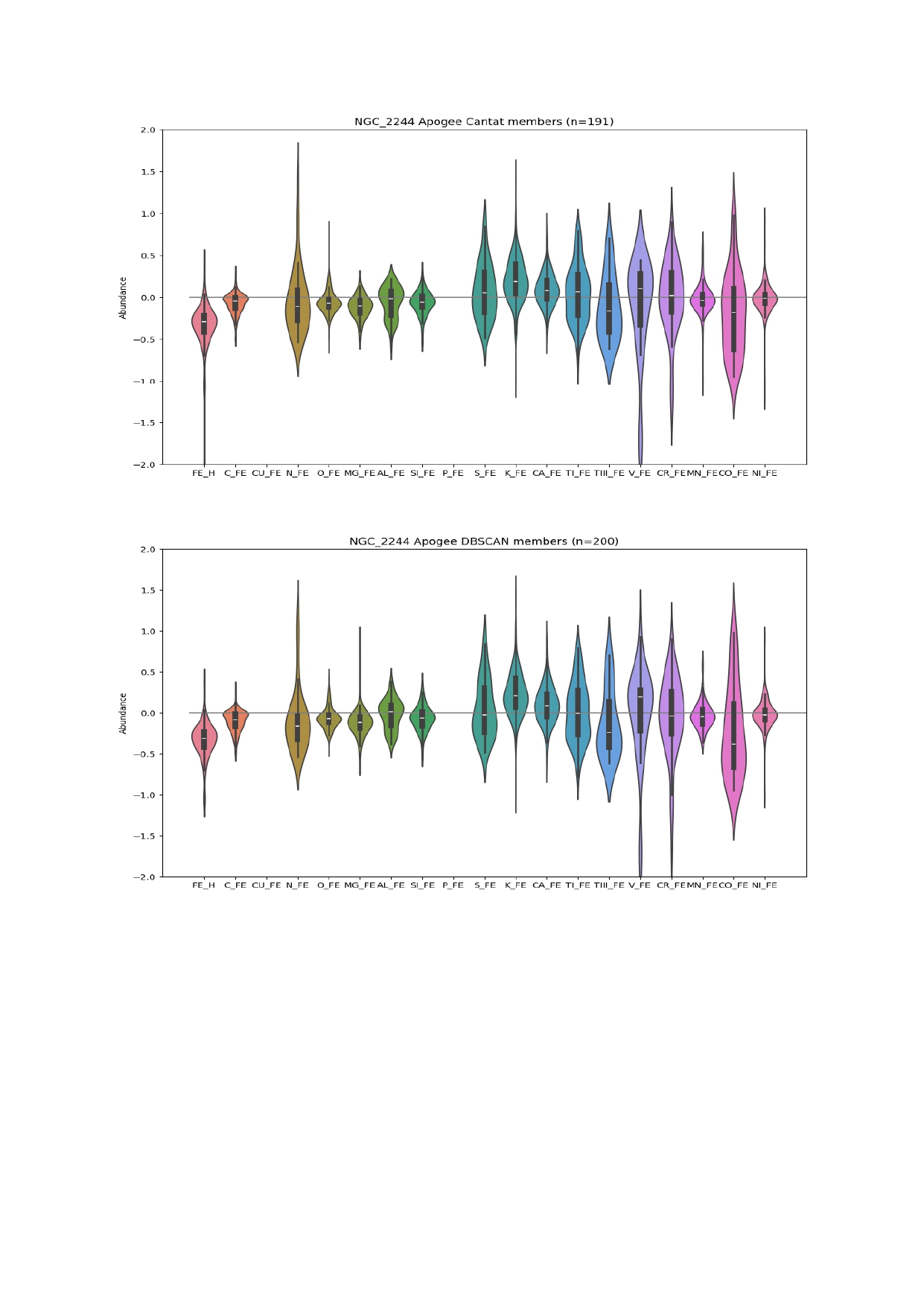}
     \caption[APOGEE data for NGC~2244]{Chemical abundances of members from APOGEE for NGC 2244 (a) Upper plot \citep{cantat2018gaia} (b) Our results
}
\label{a2244}
\end{center}
\end{figure}

\begin{figure}[h]
     \begin{center}
   \includegraphics[width=1.2\linewidth,height=1.0\linewidth]{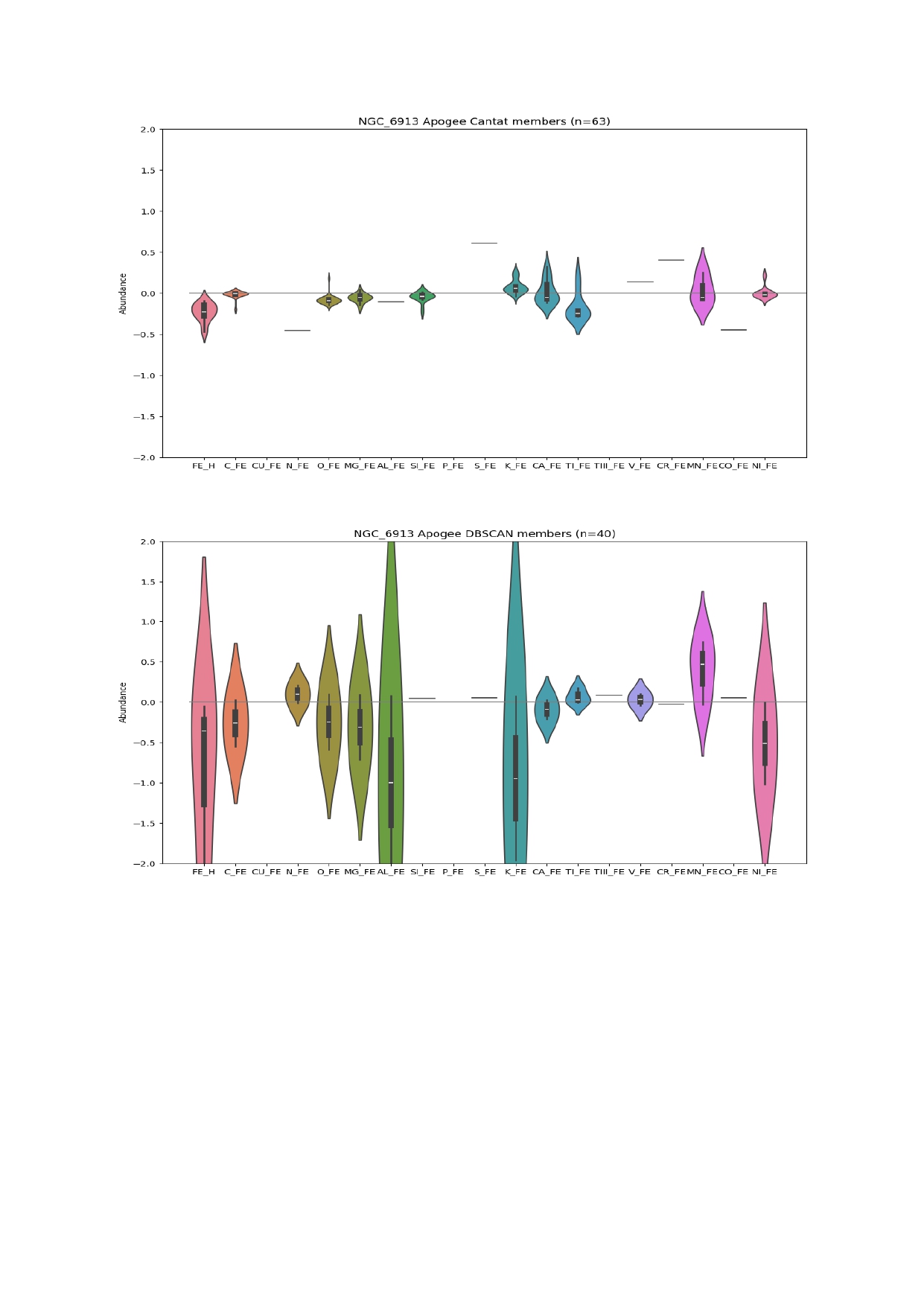}
     \caption[APOGEE data for NGC~6913]{Chemical abundances of members from APOGEE for NGC 6913 (a) Upper plot \citep{cantat2018gaia} (b) Our results
}
\label{a6913}
\end{center}
\end{figure}

\begin{figure}[h]
     \begin{center}
   \includegraphics[width=1.2\linewidth,height=1.0\linewidth]{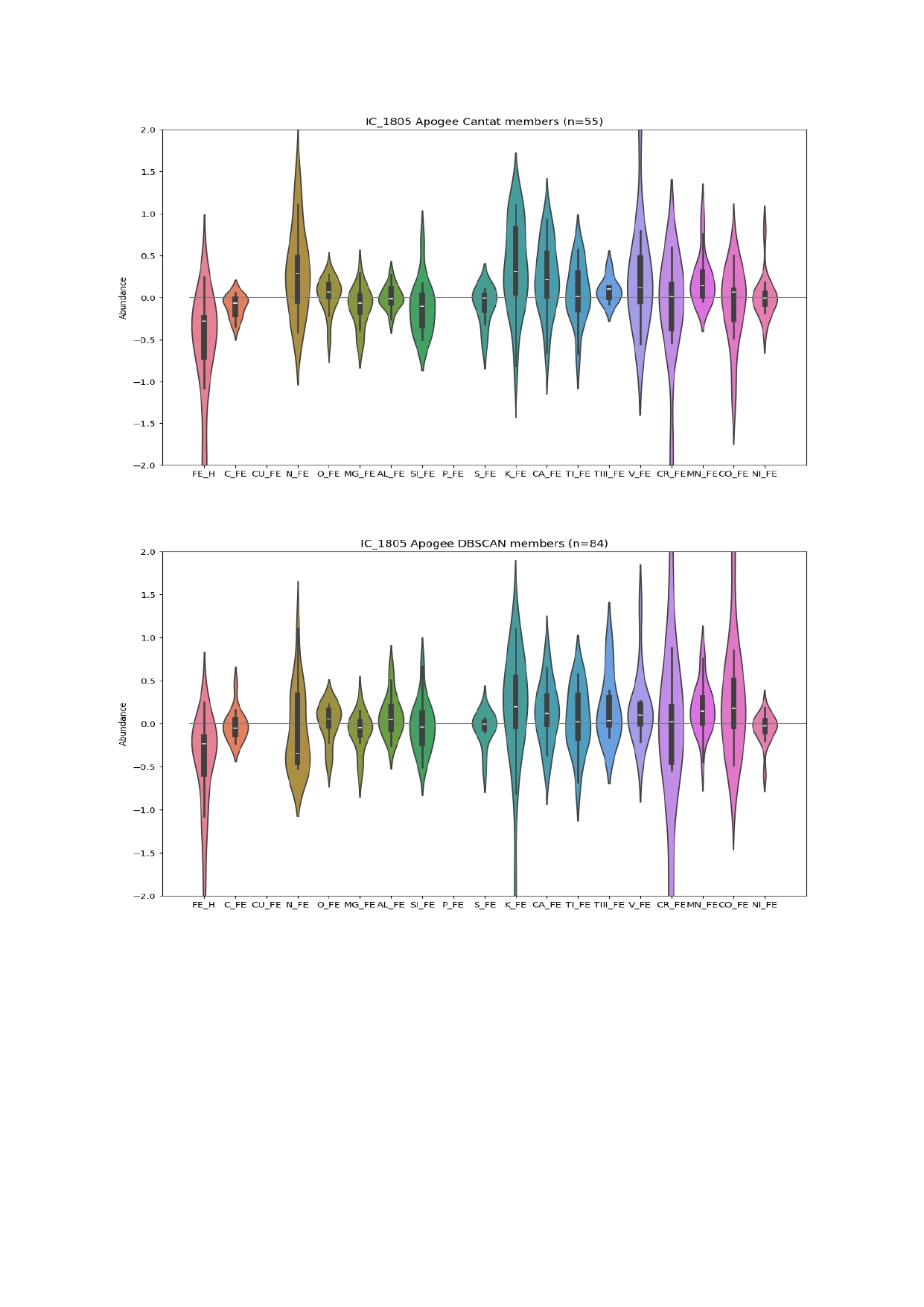}
     \caption[APOGEE data for IC~1805]{Chemical abundances of members from APOGEE for IC1805  (a) Upper plot \citep{cantat2018gaia} (b) Our results
}
\label{a1805}
\end{center}
\end{figure}

\begin{figure}[h]
     \begin{center}
   \includegraphics[width=1.2\linewidth,height=1.0\linewidth]{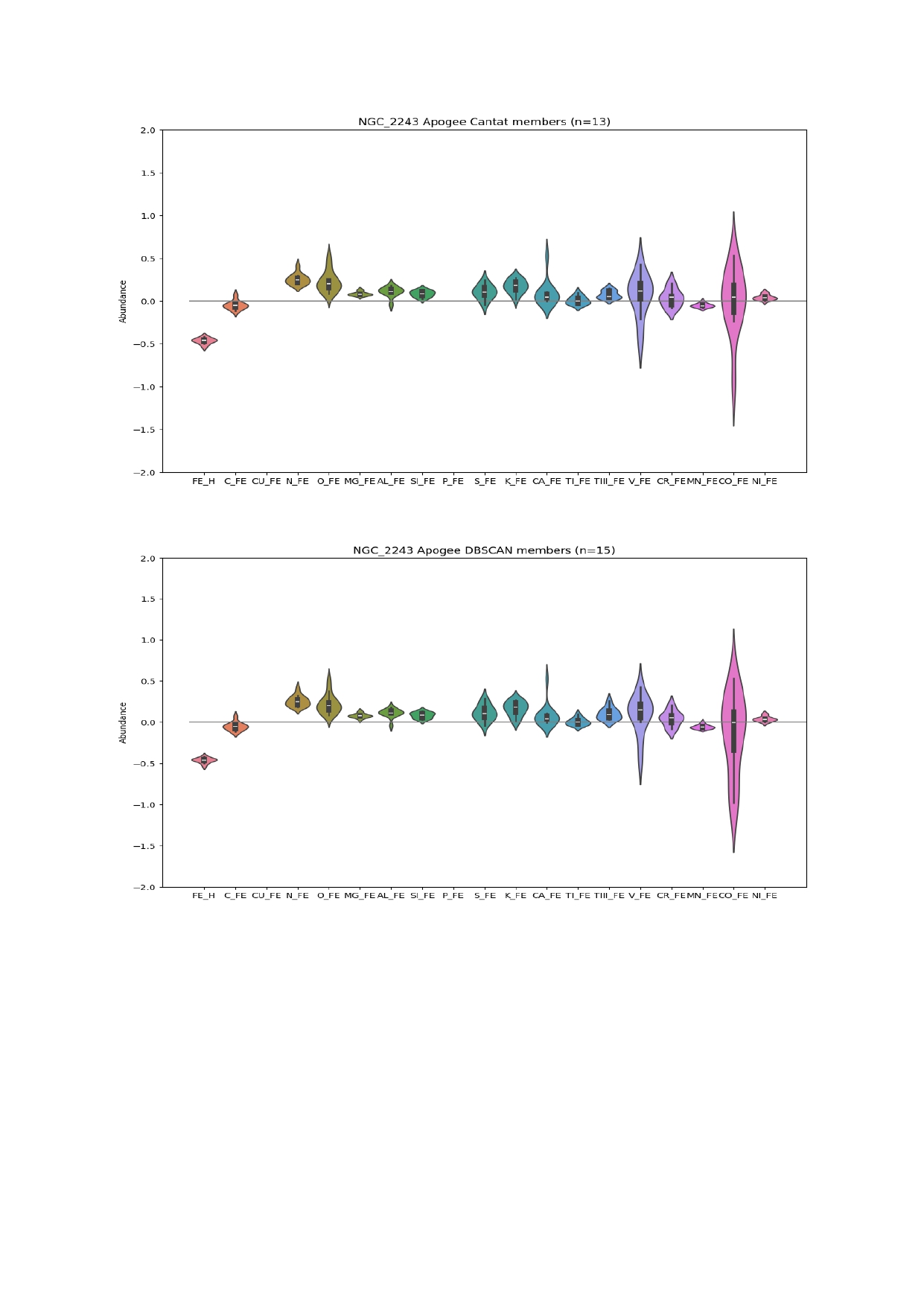}
     \caption[APOGEE data for NGC~2243]{Chemical abundances of members from APOGEE for NGC 2243 (a) Upper plot \citep{cantat2018gaia} (b) Our results
}
\label{a2243}
\end{center}
\end{figure}

Figures \ref{a2682}  to \ref{a2243} show the chemical abundances of members from APOGEE. The upper plots shows members found by \cite{cantat2018gaia} and the lower ones are our results.

\begin{figure}[h]
     \begin{center}
   \includegraphics[width=1.2\linewidth,height=1.0\linewidth]{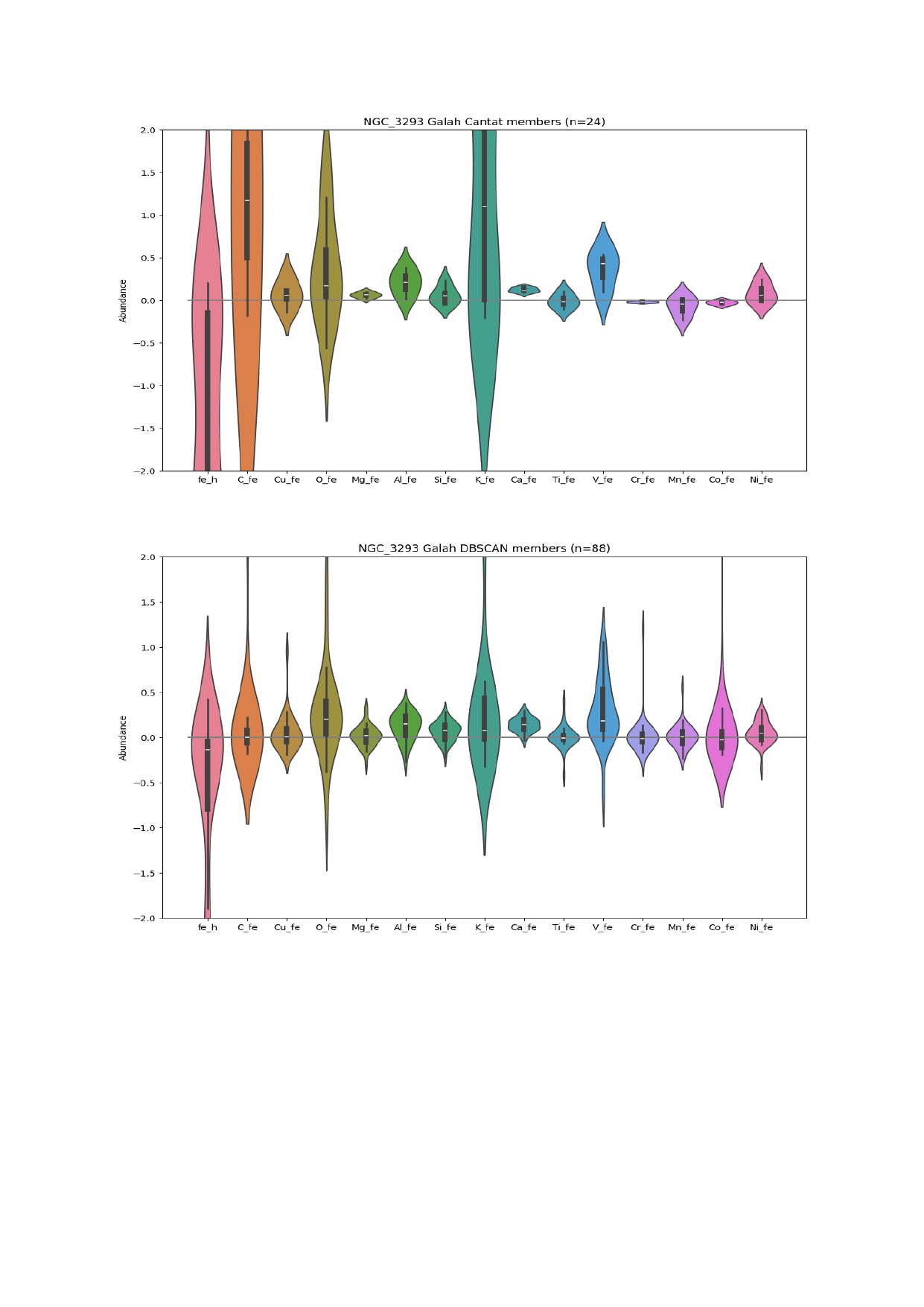}
     \caption[GALAH data for NGC~3293]{Chemical abundances of members from GALAH for NGC~3293 (a) Upper plot \citep{cantat2018gaia} (b) Our results
}
\label{G3293}
\end{center}
\end{figure}

\begin{figure}[h]
     \begin{center}
   \includegraphics[width=1.2\linewidth,height=1.0\linewidth]{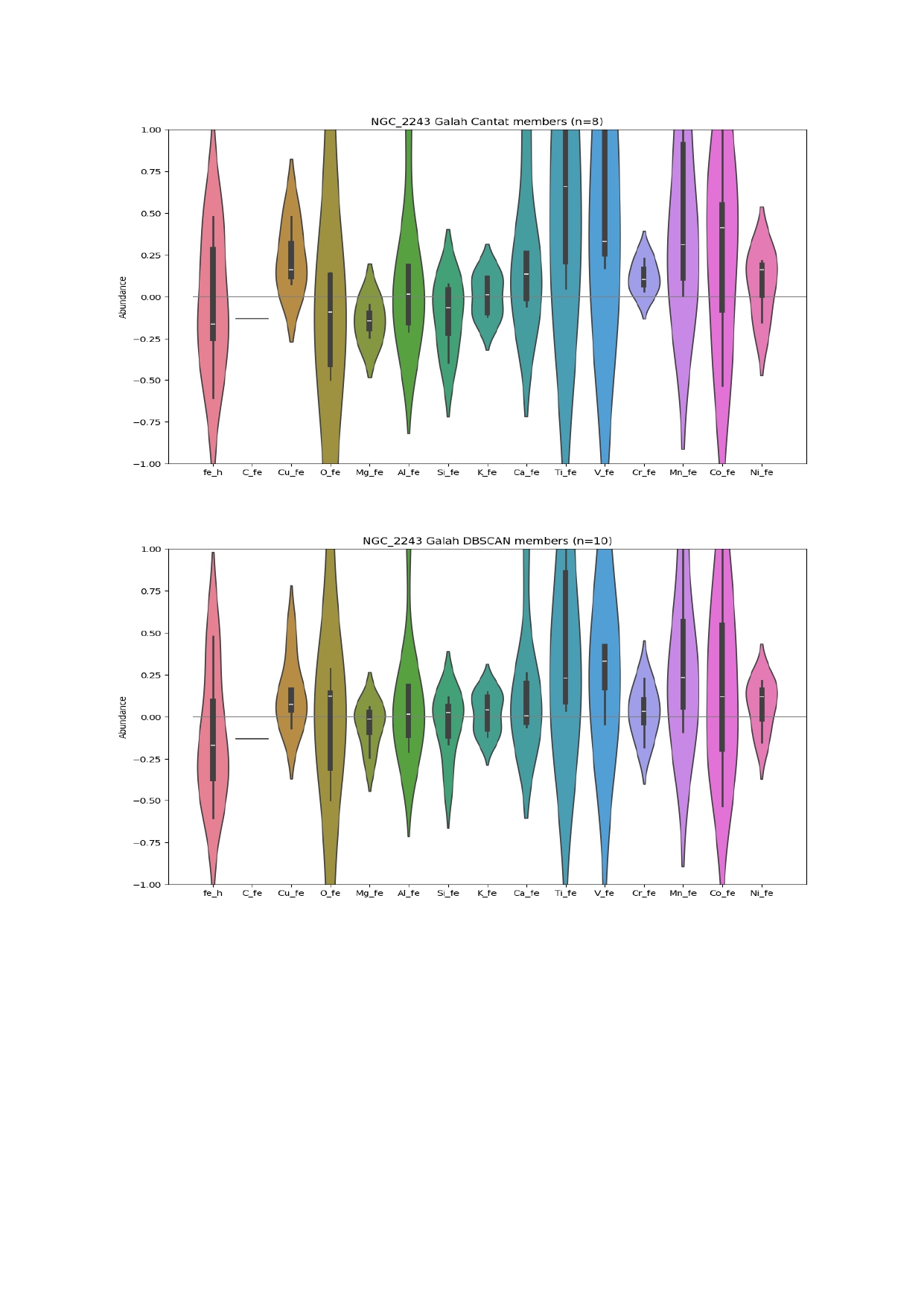}
     \caption[GALAH data for NGC~2243]{Chemical abundances of members from GALAH for NGC~2243 (a) Upper plot \citep{cantat2018gaia} (b) Our results
}
\label{G2243}
\end{center}
\end{figure}

\begin{figure}[h]
     \begin{center}
   \includegraphics[width=1.2\linewidth,height=1.0\linewidth]{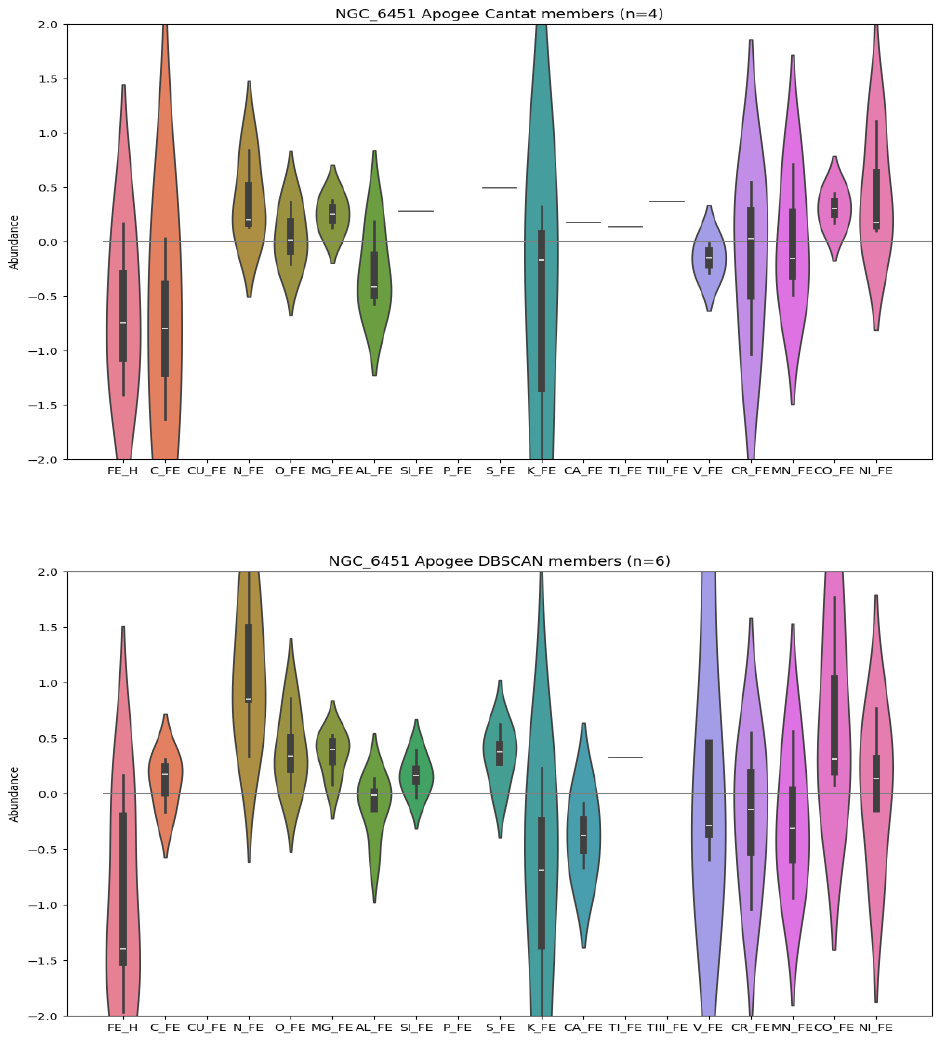}
     \caption[GALAH data for NGC~6451]{Chemical abundances of members from GALAH for NGC~6451 (a) Upper plot \citep{cantat2018gaia} (b) Our results
}
\label{G6451}
\end{center}
\end{figure}

\begin{figure}[h]
     \begin{center}
   \includegraphics[width=1.2\linewidth,height=1.0\linewidth]{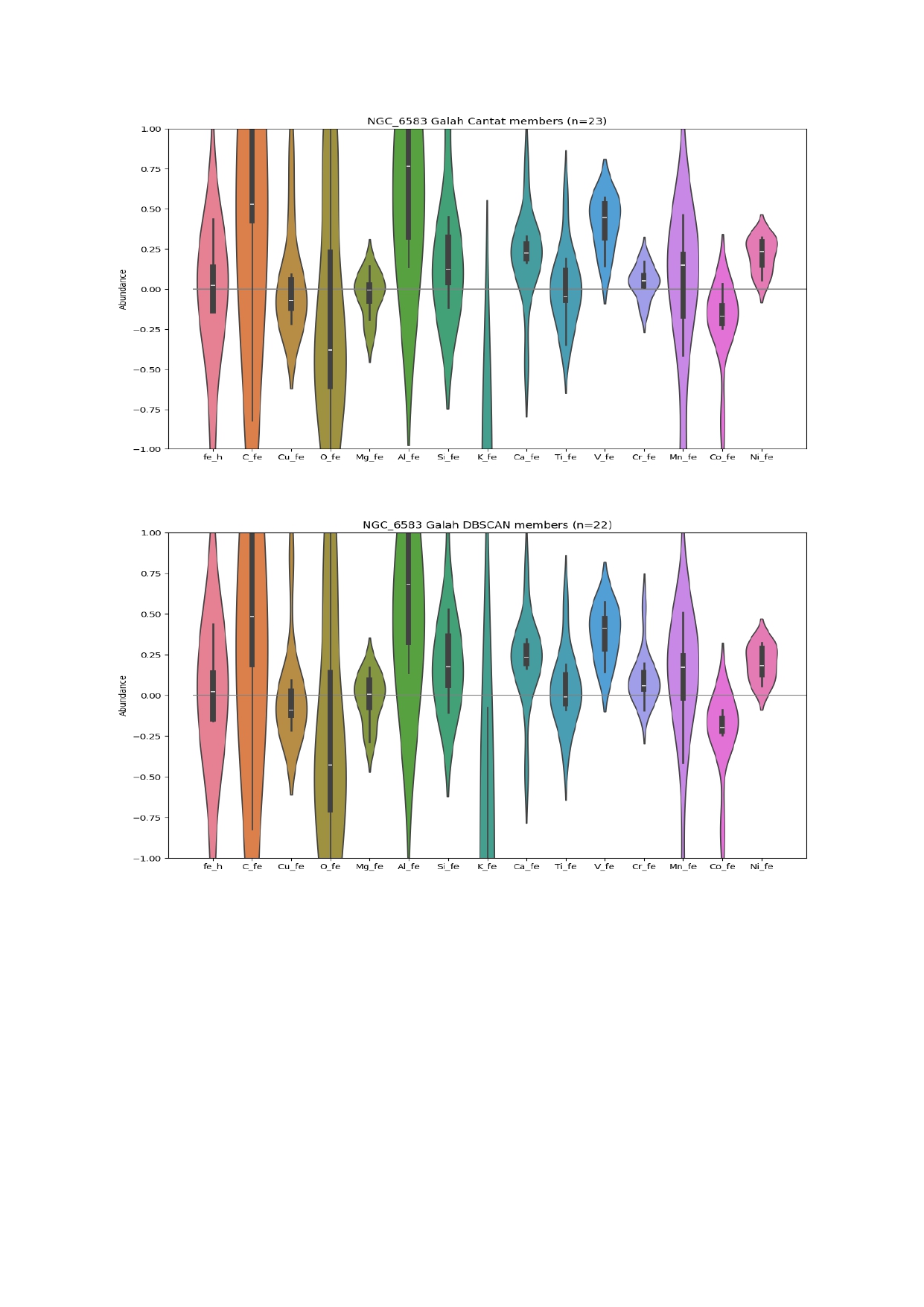}
     \caption[GALAH data for NGC~6583]{Chemical abundances of members from GALAH for NGC~6583 (a) Upper plot \citep{cantat2018gaia} (b) Our results
}
\label{G6583}
\end{center}
\end{figure}

Figures \ref{G3293} to \ref{G6583} show the chemical abundances of members from GALAH. The upper plot shows members found by \citep{cantat2018gaia} and the lower one is our result.

\section{ ASteCA plots}
\label{a2}
\begin{figure}[h]
     \begin{center}
   \includegraphics[width=0.5\textwidth]{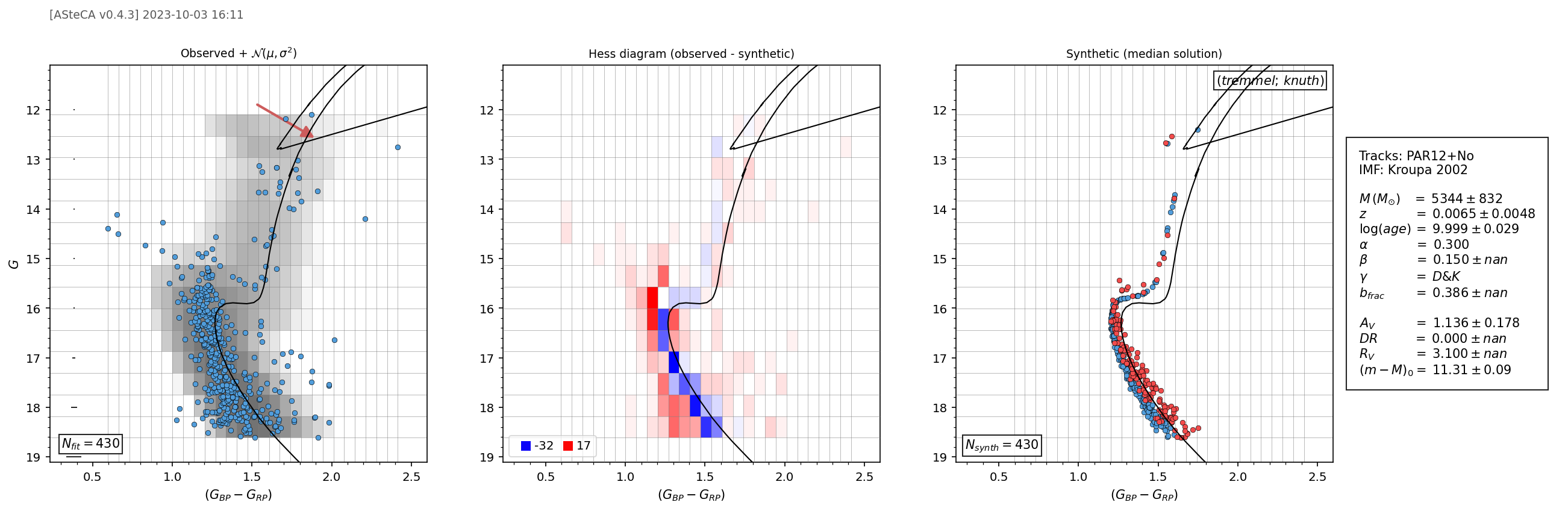}
     \caption{ASteCA CMD plots of NGC~7142}
\label{AS7142}
\end{center}
\end{figure}

\begin{figure}[h]
     \begin{center}
   \includegraphics[width=0.5\textwidth]{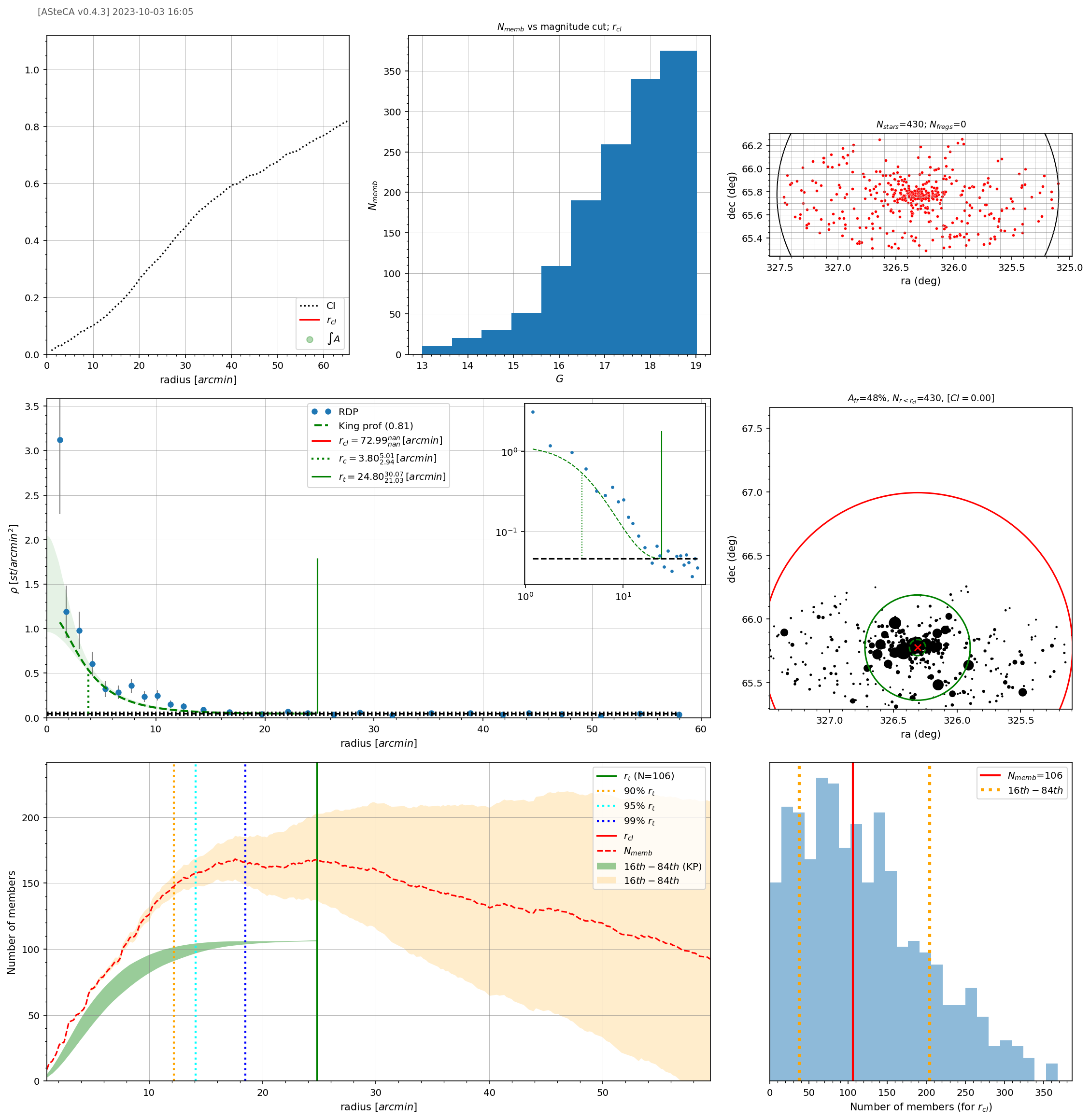}
     \caption{ASteCA plots of NGC~7142}
\label{AS7142_2}
\end{center}
\end{figure}

\begin{figure}[h]
     \begin{center}
   \includegraphics[width=0.5\textwidth]{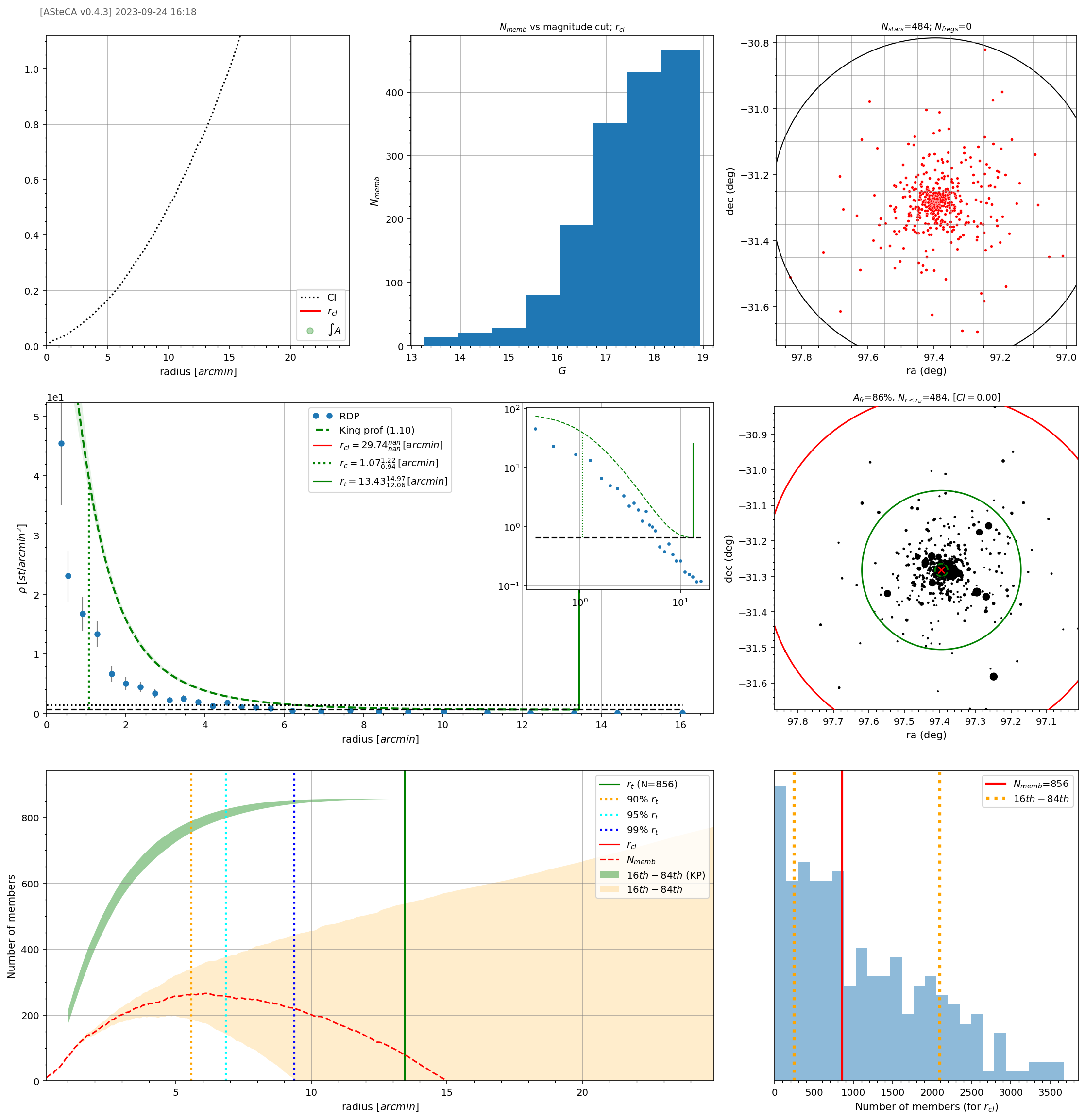}
     \caption{ASteCA plots of NGC~2243}
\label{AS2243}
\end{center}
\end{figure}

\begin{figure}[h]
     \begin{center}
   \includegraphics[width=0.5\textwidth]{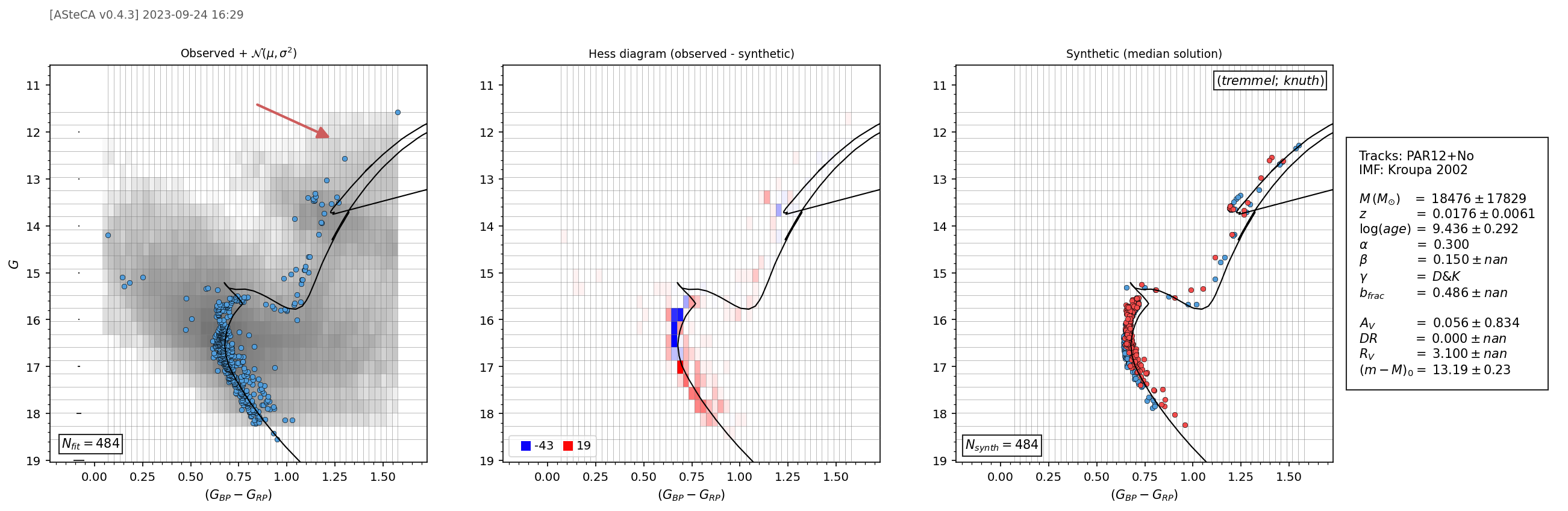}
     \caption{ASteCA CMD plots of NGC~2243}
\label{AS2243_2}
\end{center}
\end{figure}

\begin{figure}[h]
     \begin{center}
   \includegraphics[width=0.5\textwidth]{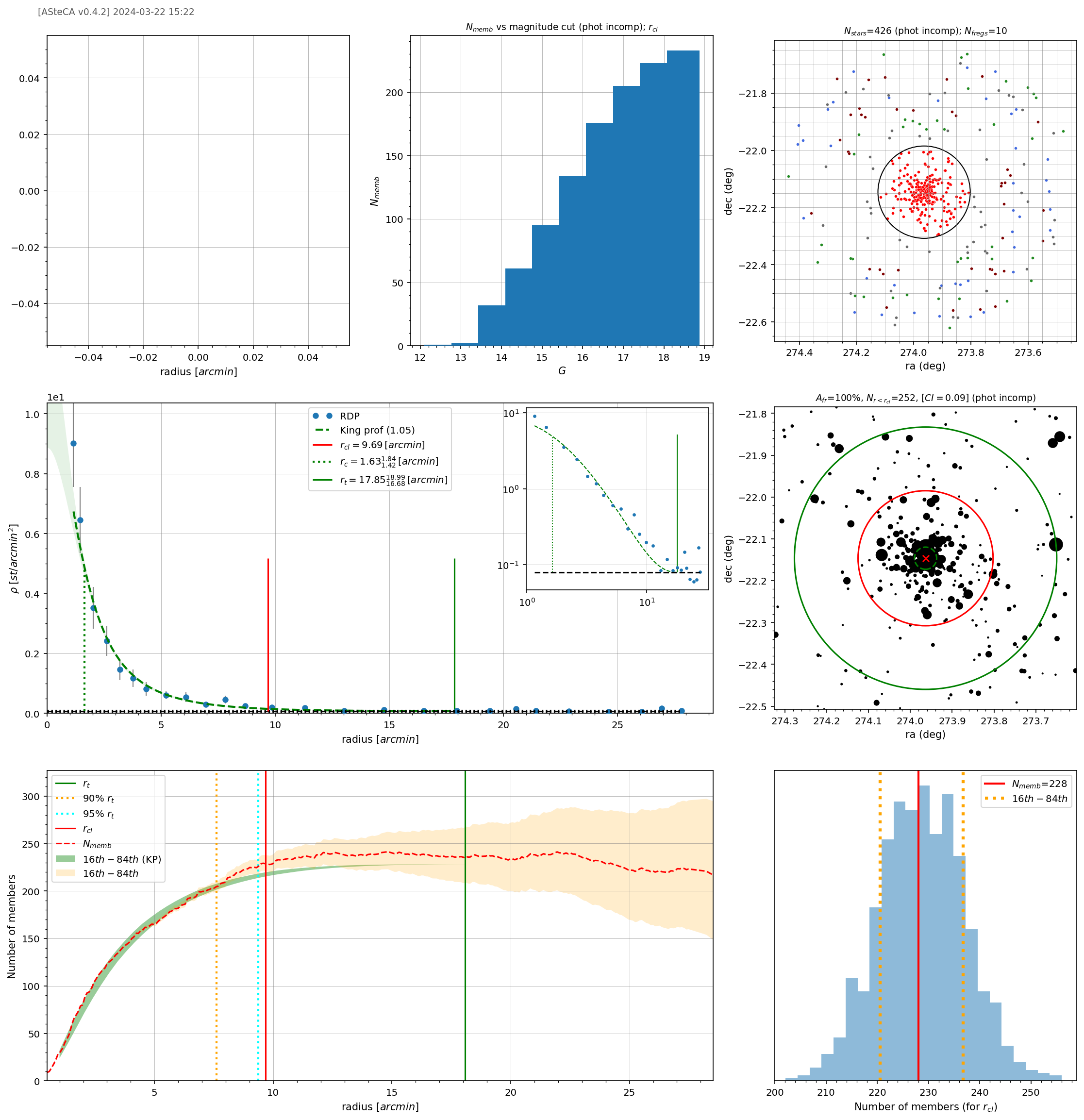}
     \caption{ASteCA plots of NGC~6451}
\label{AS6451}
\end{center}
\end{figure}

\begin{figure}[h]
     \begin{center}
   \includegraphics[width=0.5\textwidth]{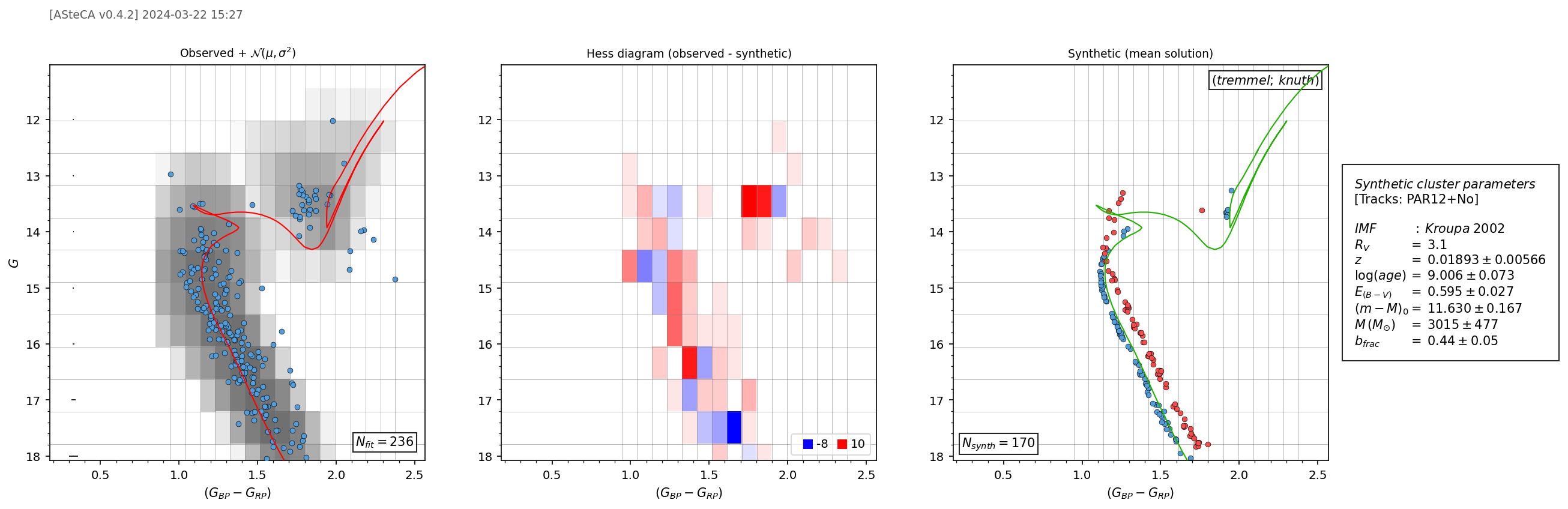}
     \caption{ASteCA CMD plots of NGC~6451}
\label{AS6451_2}
\end{center}
\end{figure}

\begin{figure}[h]
     \begin{center}
   \includegraphics[width=0.5\textwidth]{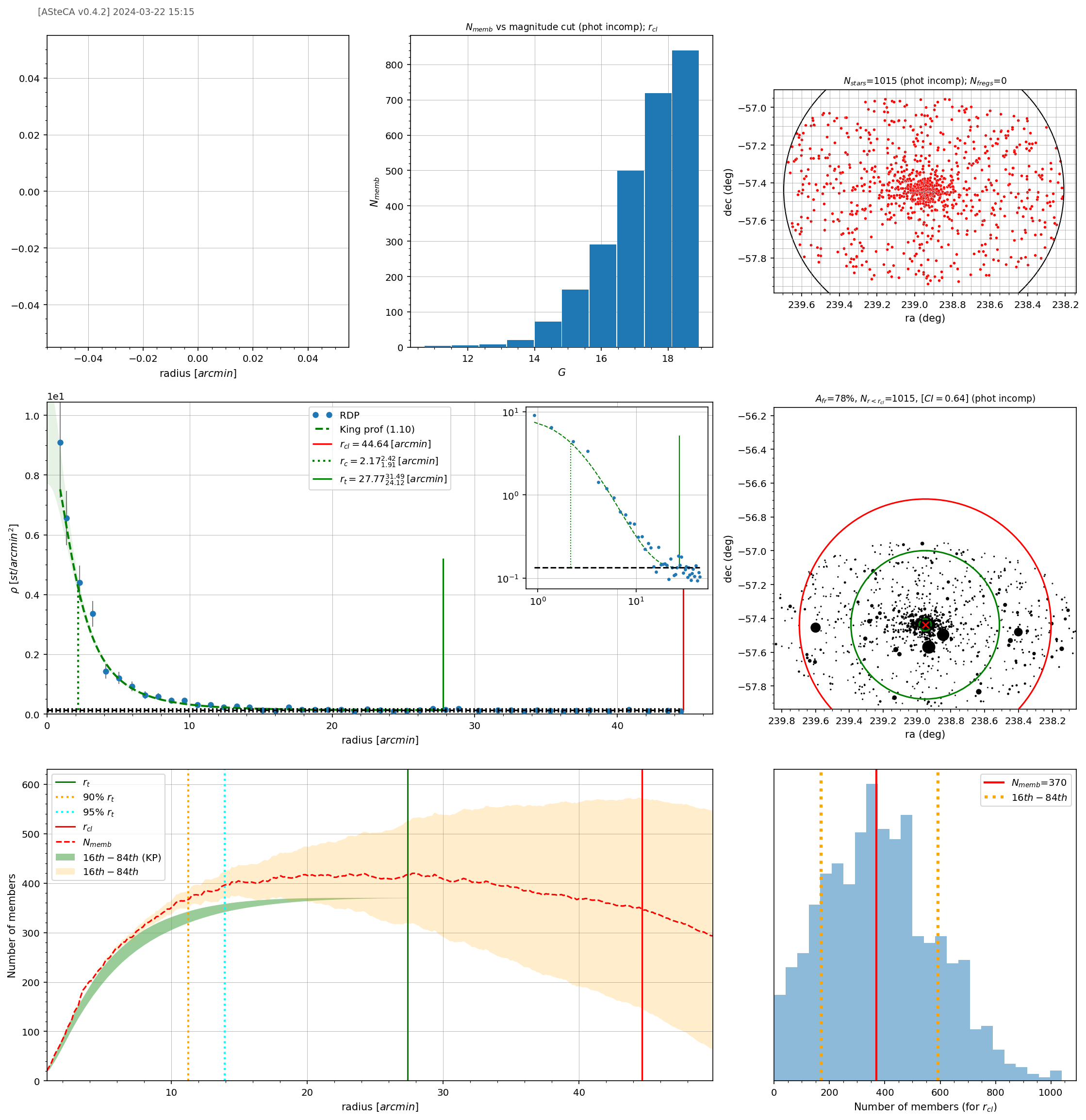}
     \caption{ASteCA plots of NGC~6005}
\label{AS6005}
\end{center}
\end{figure}

\begin{figure}[h]
     \begin{center}
   \includegraphics[width=0.5\textwidth]{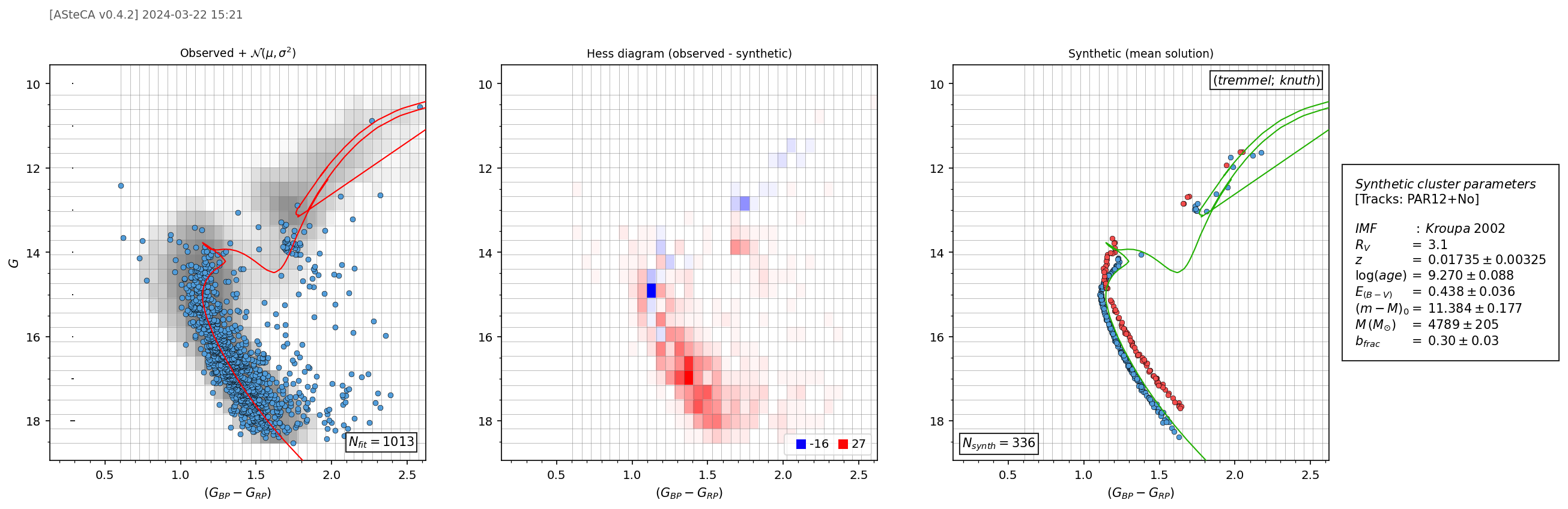}
     \caption{ASteCA CMD plots of NGC~6005}
\label{AS6005_2}
\end{center}
\end{figure}

\begin{figure}[h]
     \begin{center}
   \includegraphics[width=0.5\textwidth]{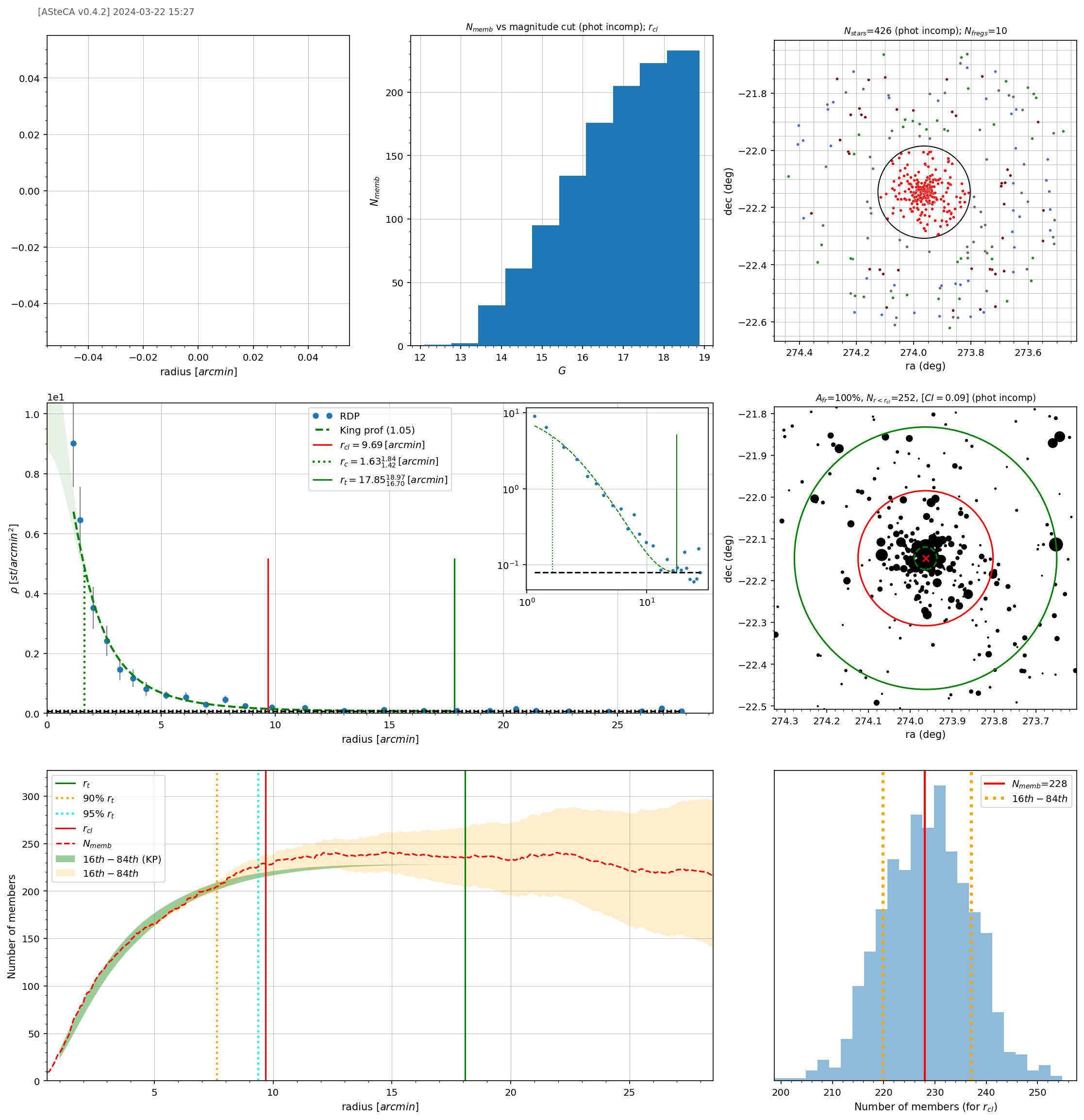}
     \caption{ASteCA plots of NGC~6583}
\label{AS6583}
\end{center}
\end{figure}

\begin{figure}[h]
     \begin{center}
   \includegraphics[width=0.5\textwidth]{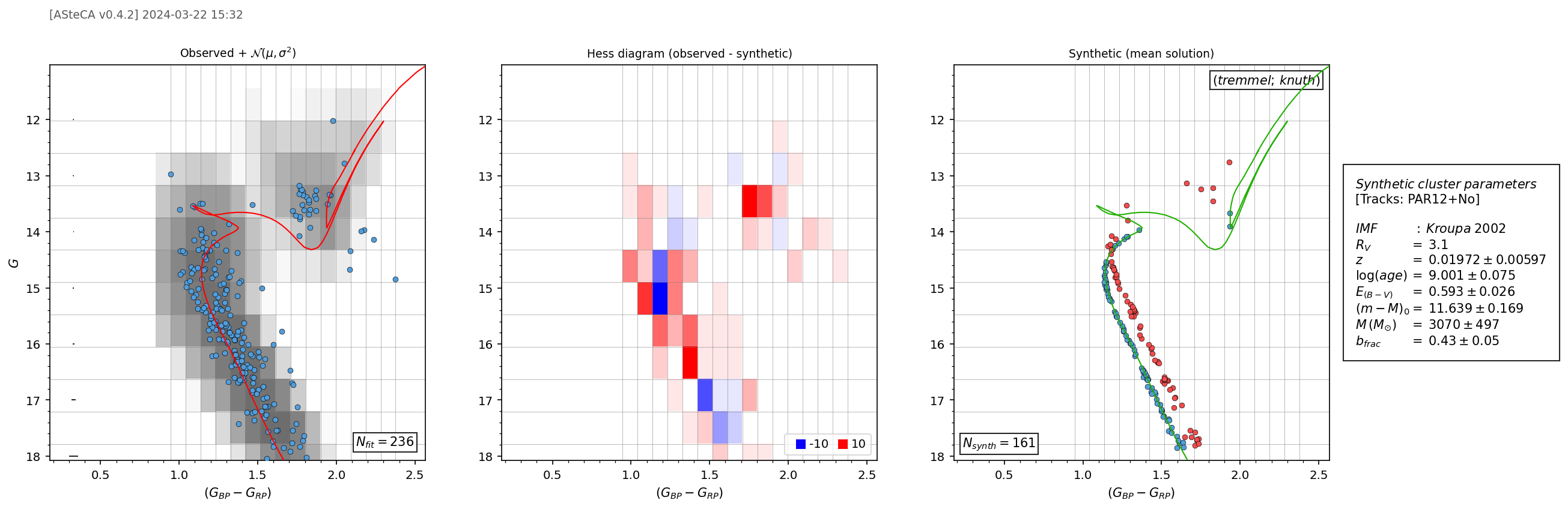}
     \caption{ASteCA CMD plots of NGC~6583}
\label{AS6583_2}
\end{center}
\end{figure}

Figures \ref{AS7142} to  \ref{AS6583_2} show  ASteCA plots  and CMDs of the revised members that we obtained after running DBSCAN clustering algorithm on our sample.

\end{document}